\documentclass[prd,aps,preprintnumbers,showpacs,superscriptaddress]{revtex4}

\usepackage{amsmath}
\usepackage{array}
\usepackage{amssymb}
\usepackage{cancel}
\usepackage{epsfig}
\usepackage{graphicx}
\usepackage{psfrag}
\usepackage{slashed}

\newcommand{\be}{\begin{equation}}
\newcommand{\ee}{\end{equation}}
\newcommand{\bea}{\begin{eqnarray}}
\newcommand{\eea}{\end{eqnarray}}

\newcommand{\bm}[1]{\mbox{\boldmath $#1$}}

\newcommand{\st}{{\scriptscriptstyle T}}
\newcommand{\sL}{{\scriptscriptstyle L}}
\DeclareMathOperator{\tr}{Tr}

\begin{document}

%
\title{Generalized universality of higher transverse moments of quark transverse momentum dependent correlators}

\author{M.G.A.~Buffing}
\email{m.g.a.buffing@vu.nl}
\affiliation{Nikhef and
Department of Physics and Astronomy, VU University Amsterdam,\\
De Boelelaan 1081, NL-1081 HV Amsterdam, the Netherlands}

\author{A.~Mukherjee}
\email{asmita@phy.iitb.ac.in}
\affiliation{Department of Physics, Indian Institute of Technology Bombay, Powai, Mumbai 400076, India}
\affiliation{Institut f\"ur Theoretische Physik, Universit\"at T\"ubingen, 72076 T\"ubingen, Germany}

\author{P.J.~Mulders}
\email{mulders@few.vu.nl}
\affiliation{Nikhef and
Department of Physics and Astronomy, VU University Amsterdam,\\
De Boelelaan 1081, NL-1081 HV Amsterdam, the Netherlands}

\begin{abstract}
The color gauge-invariant transverse momentum dependent (TMD) quark 
correlators contain process dependent gauge links in the bilocal 
matrix elements. In this paper, we split these process dependent 
correlators into universal 
TMD correlators, which in turn can be parametrized in universal TMD 
distribution functions. The process dependence is contained in 
gluonic pole factors, of which the value is determined by the
gauge link. The operator structures of the universal TMD correlators 
are identified using transverse moments.
In this paper, specific results for double transverse weighting of 
quark TMDs are given.
In particular, we show that for a spin 1/2 target one has three
universal time-reversal even leading `pretzelocity 
distributions', two of which involve double gluonic pole matrix elements and come 
with process dependent gluonic pole factors. We generalize the 
results for single and double weighting to TMD correlators of any
specific rank, illustrating it for unpolarized, spin 1/2 and spin 1 targets.
\end{abstract}
\date{\today}

\maketitle

\section{Introduction}
The transverse momentum dependent (TMD) correlators for quarks and 
gluons include not only the dependence on the longitudinal 
momentum fraction $x$ but also the dependence on the transverse 
momentum $p_{\scriptscriptstyle T}$ 
of the partons (quarks or gluons). This enables one to 
incorporate spin-momentum correlations. The correlators in turn are
parametrized in terms of parton distribution functions (PDF) and parton
fragmentation functions (PFF). The leading TMD distribution and
fragmentation functions in these
correlators include besides the well-known spin-spin densities that
survive in the collinear case (where the transverse momentum is 
integrated) also spin-orbit densities. These provide for instance a 
natural interpretation for single spin asymmetries
observed at high energies. The correlators are the nonperturbative 
objects that enter the description of high-energy scattering processes 
through a convolution with the perturbative hard scattering process. 
They constitute nonlocal matrix elements of the parton field 
operators. Collinear gluons exchanged between the soft and hard parts 
are resummed into the nonperturbative objects and show up as the 
Wilson lines or color gauge links that make the correlators 
gauge-invariant. For TMD correlators the nonlocality in the operators 
is in the transverse direction as well as longitudinal (light-like) 
direction, and there is no unique way to connect the fields 
through the gauge 
link~\cite{Collins:2002kn,Boer:2003cm,Belitsky:2002sm,Bomhof:2004aw}. 
The link depends on the 
process under consideration. In fact, in the
case of single spin asymmetries it is the closing of the gauge link
with the transverse gauge link at light-cone plus or minus infinity
that plays a major role in distinguishing time-reversal even
(T-even) and time-reversal odd (T-odd) TMD distribution and 
fragmentation 
functions~\cite{Brodsky:2002cx,Brodsky:2002rv, Bacchetta:2005rm}. 
To study the $p_{\scriptscriptstyle T}$-dependence it is convenient to
look at the transverse moments, obtained by weighting the TMD functions 
with one or more powers of 
$p_{\scriptscriptstyle T}$~\cite{Boer:2003cm,Bomhof:2007xt}. 
It has been shown in Ref.~\cite{Boer:2003cm} that single weighted 
correlators relevant for different azimuthal asymmetries can be 
expressed in terms of two collinear correlators, the first one 
containing a T-even operator combination and the second one
containing a T-odd combination. The latter involves a
quark-quark-gluon 
matrix element with vanishing gluon momentum and is known as the Efremov-Teryaev-Qiu-Sterman (ETQS)
or gluonic pole matrix 
element~\cite{Efremov:1981sh,Efremov:1984ip,Qiu:1991pp,Qiu:1991wg,Qiu:1998ia,Kanazawa:2000nz}. 
This matrix element appears 
in cross sections multiplied with a process dependent gluonic pole 
factor, which depends on the hard part of the process. Examples of such
process dependent T-odd functions are the Sivers and Boer-Mulders 
function. The $p_{\scriptscriptstyle T}$-weighted 
moment of the fragmentation correlators can also be divided into two 
parts similar to distributions, but here T-odd effects can come also 
from the fact that one has complex non-plane wave final 
states~\cite{Boer:2003cm}. For fragmentation 
the gluonic pole matrix elements vanish and since for a given 
transverse moment 
there is only one specific operator combination, there is no process 
dependence~\cite{Collins:2004nx,Gamberg:2008yt,Meissner:2008yf,Gamberg:2010gp,Metz:2002iz}. 
This is for example the situation for the Collins fragmentation 
function. Generally, the dependence on the gauge link complicates the universality properties of TMDs as well as factorization issues. 

In terms of transverse moments, the study of TMD correlators becomes 
simpler. There remains a process dependence, but this is
dealt with by process dependent gluonic pole factors, that depend
on the hard part of the process. 
In other words, given a process one already 
knows which correlators are important. 
While single weighted moments are important for $\cos(\varphi)$ and 
$\sin(\varphi)$ asymmetries,
one needs higher $p_{\scriptscriptstyle T}$-moments for 
$\cos(n\varphi)$ or $\sin(n\varphi)$
asymmetries. For double weighted transverse moments, one looks at 
weighting with 
$p_{\scriptscriptstyle T}^\alpha p_{\scriptscriptstyle T}^\beta$. 
The pretzelocity TMD PDF $h_{1T}^\perp(x,p_T^2)$ is an example for 
which double
weighting is important. It has received some attention in the 
literature recently. 
It is a twist two chiral odd and T-even TMD distribution function. It 
contributes to the $\mathrm{sin}(3 \varphi-\varphi_S)$ asymmetry in semi-inclusive deep inelastic 
scattering (SIDIS)~\cite{Mulders:1995dh,Avakian:2008ae} and to the 
$\mathrm{cos}(2 \varphi +\varphi_a-\varphi_b)$ asymmetry 
in the Drell-Yan process involving two transversely polarized
protons~\cite{Tangerman:1994eh,Zhu:2011zm}. In some models, such as the bag model 
and spectator model, the pretzelocity distribution is shown to be 
related to the 
difference between the helicity distribution and the transversity 
distribution of 
the nucleon~\cite{Avakian:2008ae}. This relation is not expected 
to hold in the presence of gluonic interactions. In this work, we 
analyze the double $p_\st$-moment 
of quark correlators taking into account the gauge link, and show that 
like the first moment these 
can also be separated into a T-even and a T-odd part. The T-even part 
contains three contributions, two of them coming from 
quark-quark-gluon-gluon matrix elements containing two zero momentum 
gluons, which are double gluonic pole matrix elements. 
The coefficients of these matrix elements depend on the gauge link $U$ 
and are process dependent, showing that also the T-even pretzelocity
PDF $h_{1T}^{\perp [U]} (x,p_T^2)$ is nonuniversal.
We will show that the pretzelocity function is a combination of 
universal functions, linked to the three possible T-even matrix elements. 
While these three functions themselves are by construction universal, 
it is a particular combination that appears in a given process with link dependence in the multiplicative coefficients. 
The appearance of three pretzelocity functions is a striking 
example of how the separation of the correlator into T-even and T-odd 
contributions is no longer enough to isolate the process dependent part 
of the correlator when higher transverse moments are involved. 
In addition to this, we will extend the transverse moment analysis
to give definitions of universal $p_T^2$-dependent functions of
a definite rank. This will be done in general for targets with spin
and illustrated for unpolarized, spin 1/2 and spin 1 targets.

\section{Formalism}
\subsection{Starting points}
\begin{figure}[b]
\epsfig{file=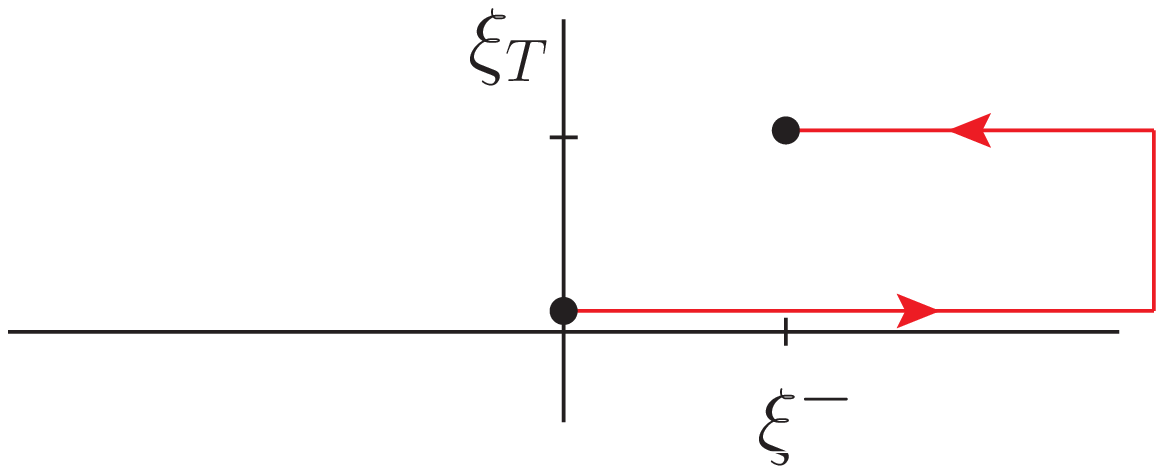,width=0.25\textwidth}
	\hspace{1.5cm}
\epsfig{file=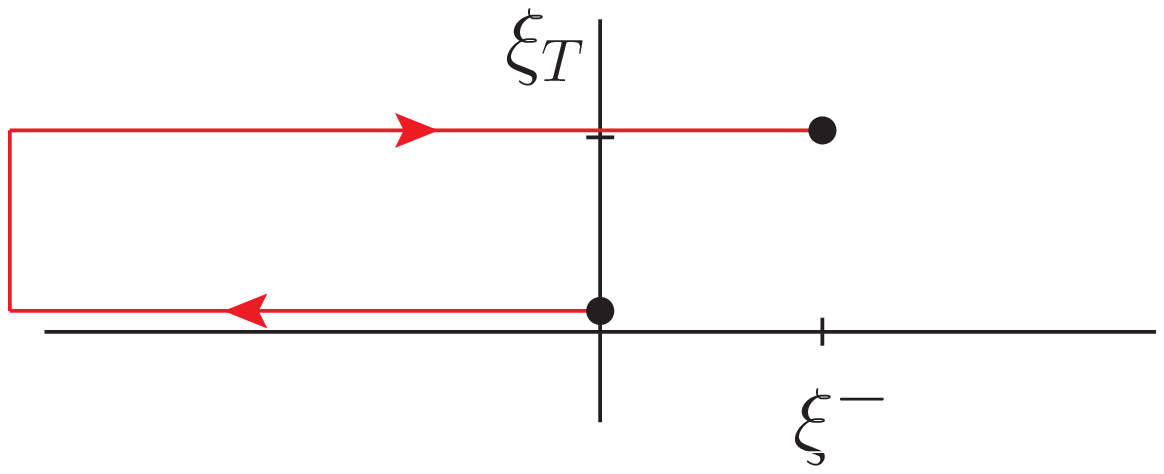,width=0.25\textwidth}
	\\[0.2cm]
	(a)\hspace{6.0cm} (b)
	\caption{\label{f:links}
	The gauge links (a) $[+]$ and (b) $[-]$ running from $0$ to 
	$\xi$ with $\xi\cdot n = 0$. The light-like separation
	$\xi^- = \xi\cdot P$ and the transverse separation $\xi_\st$
	are nonzero.}
\end{figure}

The quark-quark TMD correlator is given by
\begin{equation}
\Phi_{ij}^{[U]}(x,p_{\st};n)
=\int \frac{d\,\xi{\cdot}P\,d^{2}\xi_{\st}}{(2\pi)^{3}}
\,e^{ip\cdot \xi} \langle P\vert\overline{\psi}_{j}(0)
\,U_{[0,\xi]}\psi_{i}(\xi)\vert P\rangle\,\Big|_{\xi\cdot n=0},
\end{equation}
where we use the Sudakov decomposition 
$p^{\mu}=x P^{\mu}+p_{\st}^{\mu}+\sigma n^{\mu}$ for the 
momentum $p^{\mu}$ of the produced quark. In this decomposition, 
$P^{\mu}$ is the momentum of the incoming hadron, which is in essence 
the leading light-like direction, while $n$ is the conjugate
light-like direction satisfying $P\cdot n = 0$. The component 
$\sigma \propto p\cdot P$ along this direction is integrated over. 
The nonlocal matrix element $\Phi_{ij}^{[U]}(x,p_{\st};n)$ contains 
a process dependent gauge link $U_{[0,\xi]}$, connecting the two 
fields. The process dependence is in the path of the gauge link.
For the two simplest possibilities, the $[+]$ and $[-]$ gauge 
links, the gauge link runs from $0$ to $\xi$ through plus or minus 
infinity along $n$, respectively. 
This is illustrated in Fig.~\ref{f:links}.
More complicated gauge links can arise as well. We refer to 
Ref.~\cite{Bomhof:2004aw} for a detailed description of these gauge 
links. After integration over transverse momenta, one has the 
quark-quark collinear correlator,
\begin{equation}
\Phi_{ij}^{[U]}(x)
=\int \frac{d\,\xi{\cdot}P}{2\pi}e^{ip\cdot \xi} 
\langle P\vert\overline{\psi}_{j}(0)U_{[0,\xi]}^{[n]}
\psi_{i}(\xi)\vert P\rangle\,\Big|_{\xi\cdot n=0,\xi_{\st}=0},
\end{equation}
where the gauge link is reduced to a straight-line gauge link or Wilson 
line, which runs from $0$ to $\xi$ along $n$. 
Since the quark-quark correlators cannot be 
calculated directly, it is common to make a parametrization that 
contains TMD or collinear PDFs, respectively. 
In the TMD case, there are for a spin 1/2 nucleon eight leading 
contributing terms in the parametrization of the 
TMD correlator~\cite{Bacchetta:2006tn},
\begin{eqnarray}
\Phi^{[U]}(x,p_{\st};n)&=&\bigg\{
f^{[U]}_{1}(x,p_\st^2)
-f_{1T}^{\perp[U]}(x,p_\st^2)\,
\frac{\epsilon_{\st}^{\rho\sigma}p_{\st\rho}S_{\st\sigma}}{M}
+g^{[U]}_{1s}(x,p_\st)\gamma_{5}
\nonumber \\&&\mbox{}
+h^{[U]}_{1T}(x,p_\st^2)\,\gamma_5\,\slashed{S}_{\st}
+h_{1s}^{\perp [U]}(x,p_\st)\,\frac{\gamma_5\,\slashed{p}_{\st}}{M}
+ih_{1}^{\perp [U]}(x,p_\st^2)\,\frac{\slashed{p}_{\st}}{M}
\bigg\}\frac{\slashed{P}}{2},
\label{e:par}
\end{eqnarray}
with the spin vector parametrized as 
$S^\mu = S_{\sL}P^\mu + S^\mu_{\st} + M^2\,S_{\sL}n^\mu$ 
and shorthand notations for $g^{[U]}_{1s}$ and $h_{1s}^{\perp [U]}$,
\begin{eqnarray}
g^{[U]}_{1s}(x,p_T)=S_{\sL} g^{[U]}_{1L}(x,p_{\st}^2)
-\frac{p_{\st}\cdot S_{\st}}{M}g^{[U]}_{1T}(x,p_{\st}^2).
\end{eqnarray}
The TMD distribution functions in this parametrization 
depend on $x$ and $p_\st^2 = -\bm p_\st^2 = -\vert p_\st\vert^2$. 
The leading contributions in the correlator all have a $\slashed{P}$ 
factor
and are distinguished by different azimuthal behavior for
the transverse vectors such as $p_\st$ and $S_\st$. The 
correlators and the TMD distribution
functions in the parametrization also depend on the gauge link. 
Time-reversal relates the functions
in $\Phi^{[U]}$ to those in $\Phi^{[U^t]}$, where $U^t$ is the 
time-reversed gauge link, which 
means interchanging the running via light-cone plus or minus infinity. 
For the functions $f_{1T}^\perp$ and $h_1^\perp$ one has 
$f_{1T}^{\perp [U]} = - f_{1T}^{\perp [U^t]}$, a property that 
is referred to as
{\em naive} T-odd. Each TMD has either zero, one or two factors of
$p_T$ as a prefactor. This will play a role when integrations 
over transverse momenta are considered. It is actually useful
to use in the parametrization irreducible (symmetric and traceless) 
tensors in the transverse space, 
\begin{equation}
p_\st^\alpha, \quad p_\st^{\alpha\beta} = p_\st^\alpha\,p_\st^{\beta} 
- \frac{1}{2}p_\st^2\,g_\st^{\alpha\beta}, \ldots \, .
\end{equation} 
Just integrating (without weights) Eq.~\ref{e:par} over 
transverse momenta, only the contributions without prefactor 
of $p_{\st}$ or traces survive, yielding
\begin{equation}
\Phi(x)=\bigg\{f_1(x) +S_{\sL}g_1(x)\gamma_{5}
+h_1(x)\,\gamma_5\,\slashed{S}_{\st}\bigg\}\frac{\slashed{P}}{2}
\label{collinear}
\end{equation}
at the leading twist two level. Here $g_1(x)$ is the integrated
version of $g^{[U]}_{1L}(x,p_\st^2)$ and $h_1(x)$ is the 
$p_\st$-integrated version of 
$h_1^{[U]}(x,p_\st^2) = h_{1T}^{[U]}(x,p_\st^2) 
+ h_{1T}^{\perp [U](1)}(x,p_\st^2)$ including a trace term, 
which involves
functions weighted with powers of $-p_\st^2/2M^2 = \bm p_\st^2/2M^2$, in general
\be
f_{\ldots}^{(n)}(x,p_\st^2) 
= \left(\frac{-p_\st^2}{2M^2}\right)^n\,f_{\ldots}(x,p_\st^2).
\label{transversemoments}
\ee
The integrated functions $f_{\ldots}^{(n)}(x)$ are usually referred to 
as transverse moments, but we will extend this name to
azimuthally averaged functions that still depend on $p_\st^2$.
The collinear PDFs in Eq.~\ref{collinear} are independent of 
the gauge link $U$. In other words, all operator definitions of these 
collinear PDFs have a unique straight-line gauge link.

The behavior of (TMD) PDFs under time-reversal can be studied. 
The functions $f_{1T}^{\perp}$ and $h_{1}^{\perp}$ are time-reversal 
odd (T-odd), while the remaining six functions are time-reversal even 
(T-even). Similarly, one can look at the behavior of the matrix 
element(s) under time-reversal. Using
the fact that the simplest gauge links for quark correlators, the 
$[+]$ and $[-]$ gauge links, 
are a time-reversal couple, one can construct T-even and T-odd TMD 
correlators~\cite{Boer:2003cm},
\begin{subequations}
\label{e:DEFINITE}
\begin{gather}
\Phi^{(\text{T-even})}(x,p_\st)
=\tfrac{1}{2}\big(\,\Phi^{[+]}(x,p_\st)\,{+}\,\Phi^{[-]}(x,p_\st)
\,\big)\ ,
\label{DEFINITE-1}\\
\Phi^{(\text{T-odd})}(x,p_\st)
=\tfrac{1}{2}\big(\,\Phi^{[+]}(x,p_\st)\,{-}\,\Phi^{[-]}(x,p_\st)
\,\big)\ .
\label{DEFINITE-2}
\end{gather}
\end{subequations}
For the unweighted integrated case the separation between T-even 
and T-odd objects would 
be trivial, since the $[+]$ and $[-]$ gauge links are identical 
after integration over transverse 
momentum. As a result, $\Phi(x) = \Phi^{(\text{T-even})}(x)$ and 
$\Phi^{(\text{T-odd})}(x) = 0$. 
For the transverse momentum weighted case both functions are 
important. One thus is tempted
to identify the TMD functions $f_{1T}^\perp$ and $h_1^\perp$ to 
the T-odd correlator and the
other TMD functions to the T-even correlator, in which the 
T-odd ones acquire process dependence.
The situation will turn out to be more complex, which is most
easily demonstrated by looking at transverse momentum weighting.

\subsection{Single transverse weighting}
\begin{figure}[bt]
	\epsfig{file=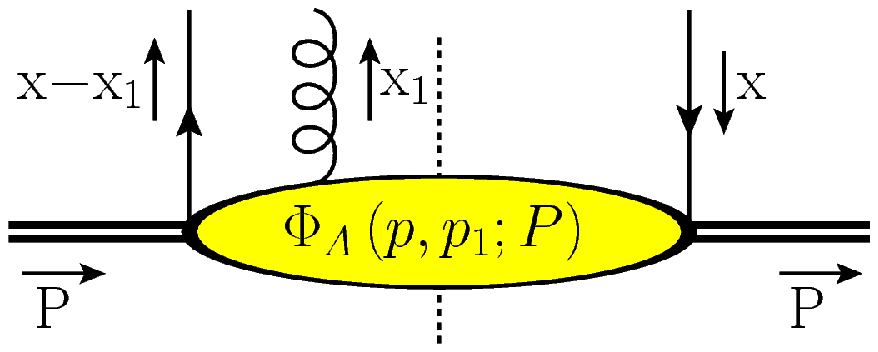,width=0.25\textwidth}
	\hspace{1.5cm}
	\epsfig{file=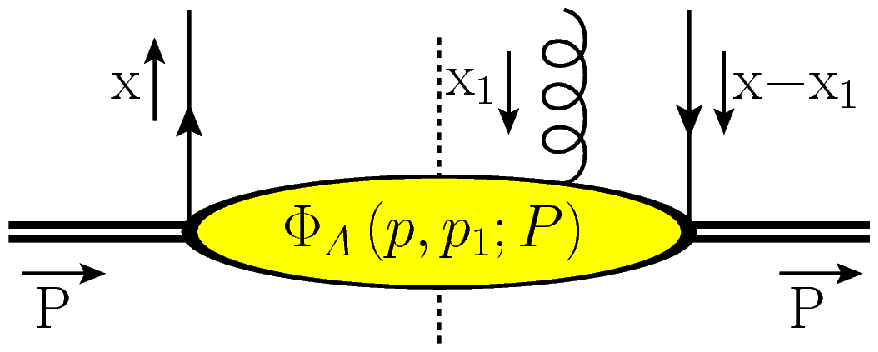,width=0.25\textwidth}
	\hspace{1.5cm}
	\epsfig{file=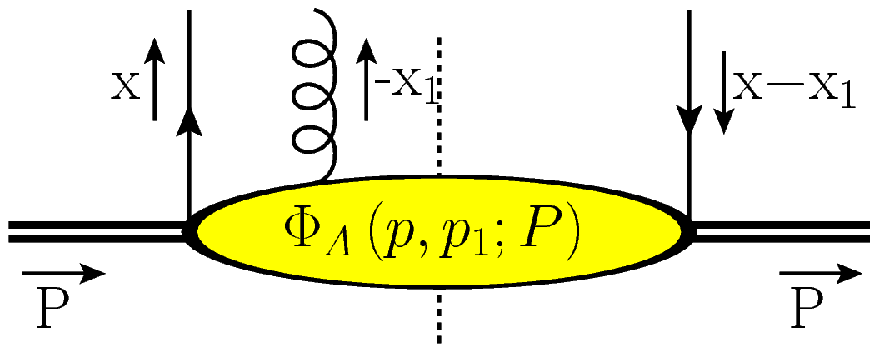,width=0.25\textwidth}
	\\[0.2cm]
	(a)\hspace{6.0cm} (b)\hspace{6.0cm} (c)
\caption{\label{f:Phi_Acor}
The correlators (a) $\Phi_{A}(x-x_{1},x_{1}|x)$, (b) $\Phi_{A}(x|x_{1},x-x_{1})$ and (c) $\Phi_{A}(x,-x_{1}|x-x_{1})$. Note that the diagrams in (b) and (c) are equal to each other.} 
\end{figure}

In the $p_{\st}$-weighted case multiple matrix elements appear, 
since the transverse weighting gives rise to a derivative that 
not only acts on the fields, but on the gauge links as well. 
Weighting with $p_\st^\alpha$ can be rewritten in terms of
two contributions, which 
upon $p_\st$-integration only depend on $x$ and depend on the link 
just through a gluonic pole factor~\cite{Boer:2003cm},
\begin{eqnarray}
\Phi_{\partial}^{\alpha[U]}(x) \equiv 
\int d^2 p_{\st}\,p_{\st}^{\alpha}\Phi^{[U]}(x,p_{\st})
&=& \Big(\Phi_{D}^{\alpha}(x)-\Phi_{A}^{\alpha}(x)\Big)
+\pi C_{G}^{[U]}\Phi_{G}^{\alpha}(x) \nonumber \\
&=&\widetilde\Phi_{\partial}^{\alpha}(x)
+\pi C_{G}^{[U]}\Phi_{G}^{\alpha}(x).
\label{e:Phip}
\end{eqnarray}
The matrix element $\Phi_{G}^{\alpha}(x)$ is referred to as the gluonic pole matrix element. All matrix elements are built from
multi-parton twist three operator combinations, illustrated
in Fig.~\ref{f:Phi_Acor}.
The relevant ones in the final result of a calculation involve
gauge-invariant operators $iD_\st^\alpha$ and $F_\st^{n\alpha}$
rather than $A_\st^\alpha$,
\bea
\Phi^\alpha_{D\,ij}(x-x_1,x_1\vert x) &=&
\int \frac{d\xi{\cdot}P\,d\eta{\cdot}P}{(2\pi)^2}
\,e^{ip_1{\cdot}\eta+i(p-p_1)\cdot \xi} 
\langle P\vert\overline{\psi}_{j}(0)
U_{[0,\eta]}\,iD_\st^\alpha(\eta)\,U_{[\eta,\xi]}
\psi_{i}(\xi)\vert P\rangle\,\Big|_{LC},
\label{e:PhiD}
\\ 
\Phi^{\alpha}_{F\,ij}(x-x_1,x_1\vert x) &=&
\int \frac{d\xi{\cdot}P\,d\eta{\cdot}P}{(2\pi)^2}
\,e^{ip_1{\cdot}\eta+i(p-p_1)\cdot \xi} 
\langle P\vert\overline{\psi}_{j}(0)
U_{[0,\eta]}\,F_\st^{n\alpha}(\eta)\,U_{[\eta,\xi]}
\psi_{i}(\xi)\vert P\rangle\,\Big|_{LC}.
\eea
The matrix elements showing up in Eq.~\ref{e:Phip} are 
related to these multi-parton correlators, to be precise
we need the matrix elements
\bea
&&
\Phi_D^\alpha(x) 
= \int dx_1\ \Phi_D^\alpha(x-x_1,x_1\vert x),
\\ &&
\Phi_{A}^{\alpha}(x)
\equiv 
\int dx_{1}\,\text{PV}\frac{i}{x_{1}}
\,\Phi_{F}^{n\alpha}(x-x_{1},x_{1}|x),
\\[0.2cm] &&
\Phi_G^\alpha(x) = \Phi^{n\alpha}_{F}(x,0|x).
\label{e:PhiG} 
\eea
By using the Eqs~\ref{DEFINITE-1} and \ref{DEFINITE-2}, one finds that 
$\widetilde\Phi_{\partial}^{\alpha}(x)$ is a T-even matrix element,
involving the $\Phi_D$ and $\Phi_A$, the latter being the principal 
value integration over a correlator involving the gluon fields 
defined in a T-invariant way.
The gluonic pole matrix element $\Phi_{G}^{\alpha}(x)$, in order to 
distinct it from $\Phi_F$ indicated with an index $G$, is T-odd. 

The choice of notation for the arguments of multiparton 
correlators, where the produced quark and gluon on the left side 
of the cut have momentum fractions $x-x_{1}$ and $x_{1}$ and the 
incoming quark on the right side of the cut has a momentum fraction 
$x$, is illustrated in Fig.~\ref{f:Phi_Acor}a. Despite the fact 
that the assignment of momenta in the correlators in Eqs~\ref{e:PhiD}-\ref{e:PhiG} 
is overdetermined, it has the advantage that it is more transparent 
in our forthcoming generalization to higher weightings. 
Furthermore, because of the absence of T-ordering, one can move a gluon through the cut by changing 
the sign of the momentum. Under hermiticity one finds that 
the correlators in Eq.~\ref{e:PhiG} have the behavior 
$\gamma_{0}\Phi_{A}^{\dagger}(x-x_{1},x_{1}|x)\gamma_{0}$
= $\Phi_{A}(x|x_{1},x-x_{1})$.

The weighting with transverse momenta can also be analyzed by studying 
the parametrization in PDFs. 
For single $p_{\st}$-weighting, only PDFs with one prefactor of 
$p_{\st}$ in the parametrization in Eq.~\ref{e:par} survive. 
T-even PDFs contribute to the $\widetilde\Phi_{\partial}^{\alpha}(x)$ 
matrix element, while T-odd PDFs contribute to the 
$\Phi_{G}^{\alpha}(x)$ matrix element, see Ref.~\cite{Boer:2003cm} for a detailed 
study of this. Since the T-odd matrix element comes with a 
process dependent prefactor, it can be seen that for single 
$p_{\st}$-weighting, the behavior under time-reversal can be used to identify the 
process dependent parts. In literature, it has become common to
use notations like $f_{1T}^{\perp\,(\text{SIDIS})}$ or 
$f_{1T}^{\perp\,(\text{DY})}$ for these functions. 
This suggests that there are many different versions of specific functions, which are obviously not universal, one for each process 
with a different gauge link. In fact there is a universal
transverse moment relating all link dependent ones
\begin{equation}
f_{1T}^{\perp(1)[U]}(x)=C_G^{[U]} f_{1T}^{\perp(1)}(x).
\label{signchange}
\end{equation}
Although the only difference for the single weighted case is just 
the numerical prefactor that for simple processes is just $+1$ or 
$-1$, we will show in the next section that for the double weighted 
case the situation becomes more complicated and one actually 
gains a lot by this different notation. But even for single
weighting there is a clear advantage using Eq.~\ref{signchange},
because it states that there is a universal function with 
calculable process (link) dependent numbers rather than an infinite 
number of somehow related functions. For some gauge links, these
numbers are shown in Table~\ref{t:gpfactors}. Here $U^{[\square]}$ is the Wilson loop $U^{[-]^{\dagger}}U^{[+]}$.

\subsection{Double transverse weighting}
In order to evaluate the double transverse weighting we need to 
consider matrix elements like
\begin{eqnarray}
\Phi_{FF}^{\alpha\beta}(x-x_1-x_2,x_1,x_2|x) &=&
\int\frac{d\,\xi{\cdot}P}{2\pi}\frac{d\,\eta{\cdot}P}{2\pi}
\frac{d\,\eta^{\prime}{\cdot}P}{2\pi}
\ e^{i x_2(\eta^{\prime}\cdot P)}\,e^{i x_1(\eta\cdot P)}
\,e^{i(x-x_1-x_2)(\xi\cdot P)} \nonumber \\
&&\hspace{3mm} \times \langle P,S|\overline{\psi}(0)
\,U^{[n]}_{[0,\eta^{\prime}]} F_{\st}^{n\alpha}(\eta^{\prime})
U^{[n]}_{[\eta^{\prime},\eta]} F_{\st}^{n\beta}(\eta)
U^{[n]}_{[\eta,\xi]}\,\psi(\xi)|P,S\rangle\Biggr|_{LC}, \label{e:qcor_FF_2}
\end{eqnarray}
among others, where \textit{LC} indicates that all transverse components 
and $n$-components of the coordinates are zero. Besides this matrix element one
needs $\Phi_{DF}$, $\Phi_{FD}$ and $\Phi_{DD}$ as well as
bilocal matrix elements, obtained by direct or principal value integrations over these matrix elements (as in the case of
single transverse momentum weighting) or gluonic pole matrix
elements, where $x_1$ or $x_2$ or both are zero. Explicitly,
the matrix elements are discussed in Appendix~\ref{A:matrix}.

The actual weighting of the gauge link dependent TMD 
correlator $\Phi^{[U]}(x,p_\st)$ gives
\begin{eqnarray}
\Phi^{\{\alpha\beta\}\,[U]}_{\partial\partial}(x) &\equiv &\int d^2 p_{\st}\,p_{\st}^{\left\{\alpha\right.}\,p_\st^{\left. \beta\right\}}\,\Phi^{[U]}(x,p_{\st}^{2}) \nonumber \\
&=& \widetilde\Phi^{\{\alpha\beta\}}_{\partial\partial}(x)
+\pi C_{G}^{[U]}\left(\widetilde\Phi_{\partial G}^{\{\alpha\beta\}}(x)+\widetilde\Phi_{G\partial}^{\{\alpha\beta\}}(x)\right)
+\sum_c \pi^2 C_{GG,c}^{[U]}\,\Phi_{GG,c}^{\{\alpha\beta\}}(x) \nonumber \\
&=& \widetilde\Phi^{\{\alpha\beta\}}_{\partial\partial}(x)
+\pi C_{G}^{[U]}\left(\widetilde\Phi_{\partial G}^{\{\alpha\beta\}}(x)+\widetilde\Phi_{G\partial}^{\{\alpha\beta\}}(x)\right)
+\pi^2 C_{GG,1}^{[U]}\,\Phi_{GG,1}^{\{\alpha\beta\}}(x)
+\pi^2 C_{GG,2}^{[U]}\,\Phi_{GG,2}^{\{\alpha\beta\}}(x).
\label{e:Phipp}
\end{eqnarray}
For the correlators containing two (or more) gluon fields like 
the one in Eq.~\ref{e:qcor_FF_2}, one must distinguish the 
different color structures for the correlator, hence a summation over the color structures $c$.
For double weighting, there are in the
double gluonic pole part two possible color structures related to the
appearance of the color traced Wilson loop $\tfrac{1}{N_c}\tr_c(U^{[\Box]})$. 
The differences between the two different correlators $\Phi_{GG,c}^{\{\alpha\beta\}}(x)$ are made explicit in Appendix~\ref{A:matrix}. 
Just as for the single weighted case in Eq.~\ref{e:Phip}, the 
structures $\widetilde\Phi_{\ldots}$ with one or more
partial derivatives denote differences between
correlators with a covariant derivative minus a correlator with a 
principal value integration,
e.g.\ $\widetilde\Phi^{\{\alpha\beta\}}_{\partial G}(x)$
= $\Phi^{\{\alpha\beta\}}_{D G}(x)-\Phi^{\{\alpha\beta\}}_{A G}(x)$.
For completeness, they are given in Appendix~\ref{A:matrix}.
Since the weighting is done with the symmetric combination, we
have symmetrized in the indices, which should not influence the
result. We also omitted the Dirac indices on the fields.
The precise form of all correlators in terms of matrix elements can be found in Appendix~\ref{A:matrix}.

\begin{table}
\begin{tabular}{r|c|c|c|}
$\mbox{}\qquad U\qquad\mbox{}$
&$\mbox{}\qquad U^{[\pm]}\qquad\mbox{}$
&$\mbox{}\quad U^{[+]}\,U^{[\Box]}\quad\mbox{}$
&$\mbox{}\ \tfrac{1}{N_c}\tr_c(U^{[\Box]})\,U^{[+]}\ \mbox{}$
\\[2pt]
\hline
$\rule{0pt}{4mm}\Phi^{[U]}$
&$\Phi^{[\pm]}$
&$\Phi^{[+\Box]}$
&$\mbox{}\quad\Phi^{[(\Box)+]}$
\\[2pt]
\hline
$C_{G}^{[U]}$&
$\pm1$&$3$&$1$
\\[1pt]
\hline
$C_{GG,1}^{[U]}$&
$1$&$9$&$1$
\\[1pt]
\hline
$C_{GG,2}^{[U]}$&
$0$&$0$&$4$
\\[1pt]
\hline
\end{tabular}\\[4mm]
\parbox{0.85\textwidth}{
\caption{
The values of the gluonic pole prefactors for some gauge links 
needed in the $p_{\st}$-weighted cases. Note that the value of
 $C_{G}^{[U]}$ is the same for single and double transverse weighting.
\label{t:gpfactors}}}
\end{table}

The only leading twist TMD PDF that contributes is the 
pretzelocity $h_{1T}^{\perp[U]}(x,p_\st^2)$. Since this function 
is T-even, its (double) transverse moment could be associated with 
both the matrix elements 
$\widetilde\Phi^{\{\alpha\beta\}}_{\partial\partial}(x)$ 
and $\pi^{2}\Phi_{GG,c}^{\{\alpha\beta\}}(x)$. 
The gluonic pole matrix elements come with gauge link dependent 
prefactors, so the pretzelocity function as it has been defined 
in literature up to now is not universal. The double gluonic 
pole factor gives the gauge link dependence and one must
identify the gauge link dependent function as the sum of
three functions,
\be
h_{1T}^{\perp(2)[U]}(x)=h_{1T}^{\perp(2)(A)}(x)
+C_{GG,1}^{[U]}h_{1T}^{\perp(2)(B1)}(x)
+C_{GG,2}^{[U]}h_{1T}^{\perp(2)(B2)}(x),
\label{threepretzel}
\ee
where the functions $h_{1T}^{\perp(2)(A)}(x)$, 
$h_{1T}^{\perp(2)(B1)}(x)$ and $h_{1T}^{\perp(2)(B2)}(x)$ are 
universal. For simple processes 
like SIDIS and Drell-Yan with $C_{GG,1}=1$ and $C_{GG,2}=0$ one finds 
just the sum of 
two pretzelocity functions. For processes with a more 
complicated gauge link structure, other combinations involving 
three functions will appear, as can be seen in Table~\ref{t:gpfactors}. 
The double-weighted results also show that for higher transverse
moments, and hence also for the full $p_\st$-dependent treatment,
a separation in T-odd and T-even functions is no longer sufficient 
to isolate the process dependent parts.

\section{Defining TMDs}
For the definition of a TMD correlator
parametrized in terms of PDFs (or PFFs) depending on the collinear
fraction $x$ and transverse momentum $p_\st$ it is important
to keep in mind the role of $x$ and $p_\st$. These are 
identified with kinematic variables in a high-energy
scattering process. This is most well-known for $x$, which
in a SIDIS process is identified with the Bjorken scaling
variable. In the same way the transverse momentum can
be identified, e.g.\ from the noncollinearity of produced
hadrons or from jet-jet asymmetries, even if for 
transverse momenta the identification is usually 
contained in a
folding of transverse momenta of several hadron correlators.
So for purposes of further analyzing we assume that we know
that a hadron correlator depends on `measurable' $x$ and 
$p_\st$. 
When including the additional collinear gluons producing the gauge link, $p_\st$ is the sum of all transverse momenta of the partons exchanged between the (soft) hadron correlator and the hard process. This requires integration over the transverse momenta of collinear gluons.

The leading relevant TMD operator structure for our
considerations thus will be of the generic bilocal form
\be 
\Phi^{[U]}(x,p_\st;n) 
= \left.\int \frac{d\xi{\cdot}P\,d^2\xi_\st}{(2\pi)^3}
\ e^{ip\cdot\xi}\,\langle P\vert \overline\psi(0)\,U_{[0,\xi]}
\,O (\xi)\,\psi(\xi)\vert P\rangle\right|_{\xi{\cdot}n=0},
\label{e:qcor_O1O2}
\ee
where $U$ is one of the possible gauge links for TMD correlators.
We can define in this way $\widetilde\Phi_\partial^\alpha(x,p_\st)$
= $\Phi_D^\alpha(x,p_\st) - \Phi_A^\alpha(x,p_\st)$,
as well as correlators
$\Phi_{DD}^{\alpha\beta}$, $\Phi_{DA}^{\alpha\beta}, \ldots,\widetilde\Phi_{\partial\partial}^{\alpha\beta}$,
$\widetilde\Phi_{\partial G}^{\alpha\beta}, \ldots,
\Phi_{GG}^{\alpha\beta}$, where one in particular for $\Phi_{GG}$
must take care of the color structure.
 
Our identification of operator structures and TMD functions in
the parametrization of correlators depends on the comparison
of moments in $x$ and $p_\st$, even if such moments in real
life are limited by kinematics of the process. This is
well-known, but nontrivial, for the collinear dependence,
where the moments can be related to local matrix elements of
quark and gluon fields. All these operators have the same 
twist (canonical dimension minus rank of Lorentz indices),
which means they contribute at the same order of the hard scale.
The $x^{N-1}$ Mellin moments correspond to
expectation values of leading twist operators of 
rank $N$. These local matrix elements have a calculable
scale dependence governed by the anomalous dimension
of the local operator. The scale is usually identified with 
the kinematic limit such as the exchanged momentum $Q^2$ in SIDIS.
The $x$-dependent functions can be reconstructed from the
Mellin moments. Their scale dependence then is obtained by
folding them with splitting functions, of which 
the Mellin moments are precisely the anomalous dimensions.

\subsection{TMDs of definite rank}

For the $p_\st$-dependent functions we follow a
similar procedure. An expansion of TMDs involves the
symmetric traceless tensors $p_\st^{\alpha_1\ldots\alpha_m}$
of rank $m$. These traceless tensors satisfy
\be 
\int d^2p_\st\ p_\st^{\alpha_1\ldots\alpha_m}
\,p_{\st i_1\ldots i_m}\,f_{\ldots}(x,p_\st^2)
\ \propto\ f_{\ldots}^{(m)}(x).
\ee
Actually, it is sufficient and for our purposes desirable to 
integrate only the azimuthal part,
\be 
\int \frac{d\varphi_p}{2\pi}\ p_\st^{\alpha_1\ldots\alpha_m}
\,p_{\st i_1\ldots i_m}\,f_{\ldots}(x,p_\st^2)
\ \propto\ f_{\ldots}^{(m)}(x,p_\st^2).
\label{e:TMDmoment}
\ee
The r.h.s. of these equations contain the transverse
moments $f^{(m)}$ defined in Eq.~\ref{transversemoments} 
as well as constant tensors without azimuthal dependence.
Writing the following parametrization,
\bea
\Phi^{[U]}(x,p_\st) &\ =\ &
\Phi(x,p_\st^2) 
+ \pi C_{G}^{[U]}\,\frac{p_{\st i}}{M}
\,\Phi_{G}^{i}(x,p_\st^2)
+ \pi^2 C_{GG,c}^{[U]}\,\frac{p_{\st ij}}{M^2}
\,\Phi_{GG,c}^{ij}(x,p_\st^2)
+ \pi^3 C_{GGG,c}^{[U]}\,\frac{p_{\st ijk}}{M^3}
\,\Phi_{GGG,c}^{ijk}(x,p_\st^2)
+\ldots
\nonumber \\ &\quad +&
\frac{p_{\st i}}{M}\,\widetilde\Phi_\partial^{i}(x,p_\st^2)
+ \pi C_{G}^{[U]}\,\frac{p_{\st ij}}{M^2}
\,\widetilde\Phi_{\{\partial G\}}^{\,ij}(x,p_\st^2)
+ \pi^2 C_{GG,c}^{[U]}\,\frac{p_{\st ijk}}{M^3}
\,\widetilde\Phi_{\{\partial GG\},c}^{\,ijk}(x,p_\st^2)+ \ldots
\nonumber \\ &\quad +&
\frac{p_{\st ij}}{M^2}
\,\widetilde\Phi_{\partial\partial}^{ij}(x,p_\st^2)
+ \pi C_{G}^{[U]}\,\frac{p_{\st ijk}}{M^3}
\,\widetilde\Phi_{\{\partial\partial G\}}^{\,ijk}(x,p_\st^2)
+ \ldots
\nonumber \\ &\quad +&
\frac{p_{\st ijk}}{M^3}
\,\widetilde\Phi_{\partial\partial\partial}^{\,ijk}(x,p_\st^2) 
+\ldots \, ,
\label{TMDstructure}
\eea
we reproduce the moments. 
Note that in Eq.~\ref{TMDstructure} depending on the gauge link there are multiple contributing color structures for terms with two or more gluonic pole terms, hence the inclusion of the summation over these color structures $c$. For the collinear correlators $\Phi_{GG,c}^{ij}(x)$ this is discussed in Appendix~\ref{A:matrix}, for the TMD 
correlator $\Phi_{GG,c}^{ij}(x,p_\st^2)$ it is discussed in
Appendix~\ref{A:matrixtransverse}. 
The operator structures on the r.h.s. 
thus are the ones appearing in an angular expansion, in which
the azimuthal dependence is made explicit. The combinations like
$\widetilde\Phi_{\{\partial G\}}$ indicate symmetrized combinations
$\widetilde\Phi_{\{\partial G\}} = \widetilde\Phi_{\partial G} + \widetilde\Phi_{G\partial}$.
Upon $\varphi$-integration only the
structure $\Phi(x,p_\st^2)$ survives. Hence, we identify
this as the {\em rank zero} TMD correlator,
\be
\Phi(x,p_\st^2) = 
\bigg\{
f_{1}(x,p_\st^2)
+S_\sL\,g_{1}(x,p_\st^2)\gamma_{5}
+h_{1}(x,p_\st^2)\gamma_5\,\slashed{S}_{\st}
\bigg\}\frac{\slashed{P}}{2},
\label{rank0}
\ee
where the TMD correlator $h_1(x,p_\st^2)$ rather than 
$h_{1T}(x,p_\st^2)$ appears (see also the remark following Eq.~\ref{collinear}). Next, we look at the weighted expressions
before $p_\st$-integration in order to explicitly identify 
further TMD functions,
\bea
\frac{p_\st^\alpha}{M}\,\Phi^{[U]}(x,p_\st) &\ =\ &
\frac{p_{\st}^\alpha}{M}\,\Phi(x,p_\st^2)
-\widetilde\Phi_{\partial}^{\alpha (1)}(x,p_\st^2)
- \pi C_{G}^{[U]}\,\Phi_G^{\alpha (1)}(x,p_\st^2)
\nonumber\\[3pt] &\quad +&
\,\frac{p^{\ \alpha}_{\st \,i}}{M^2}
\,\widetilde \Phi_\partial^i(x,p_\st^2)
+\pi C_{G}^{[U]}\,\frac{p^{\ \alpha}_{\st \,i}}{M^2}
\, \Phi_G^i(x,p_\st^2) + \ldots \, .
\label{fullfirst}
\eea
The first term is obviously a term that needs to be there
because we already identified a nonzero rank 0 TMD
correlator.
The next two terms involve the T-even and T-odd transverse 
moments of $\widetilde\Phi_\partial^{\alpha}(x,p_\st^2)$
and $\Phi_G^{\alpha}(x,p_\st^2)$, respectively,
which survive $\varphi$-integration. The
other terms in Eq.~\ref{fullfirst} contain higher rank 
tensors in $p_\st$.
Comparing the unintegrated
expression with the parametrization for a spin 1/2 target 
one thus immediately identifies in addition to the rank 0
correlator in Eq.~\ref{rank0} two {\em rank one} TMD correlators,
\bea
&&\frac{p_{\st i}}{M}\,\widetilde\Phi_\partial^{i}(x,p_\st^2) =
\bigg\{
h_{1L}^{\perp}(x,p_\st^2)\,S_\sL\frac{\gamma_5\,\slashed{p}_{\st}}{M}
-g_{1T}(x,p_\st^2)\,\frac{p_\st{\cdot}S_\st}{M}\gamma_{5}
\bigg\}\frac{\slashed{P}}{2},
\\ &&
\frac{p_{\st i}}{M}\,\Phi_{G}^{i}(x,p_\st^2)
= \frac{1}{\pi}\bigg\{
-f_{1T}^{\perp}(x,p_\st^2)\,
\frac{\epsilon_{\st}^{\rho\sigma}p_{\st\rho}S_{\st\sigma}}{M}
+ih_{1}^{\perp}(x,p_\st^2)\,\frac{\slashed{p}_{\st}}{M}
\bigg\}\frac{\slashed{P}}{2},
\eea
the first one being T-even, the second one T-odd.

Before going to double weighting, it is useful to realize
that in Eq.~\ref{TMDstructure} one does not need to subtract
trace terms from the second rank correlators. This is
automatic because of the use of the tensor 
$p_\st^{\alpha\beta}$. We can write
\be 
\frac{p_{\st ij}}{M^2}\widetilde\Phi_{\{\partial G\}}^{\,ij} 
= \left(\frac{p_{\st i} p_{\st j}}{M^2} 
- \frac{p_\st^2}{2M^2}\,g_{\st ij}\right)
\widetilde \Phi_{\{\partial G\}}^{\,ij}
= \frac{p_{\st i}p_{\st j}}{M^2}\,\widetilde\Phi_{\{\partial G\}}^{\,ij}
+\widetilde \Phi_{\{\partial{\cdot}G\}}^{(1)} =
\frac{p_{\st i} p_{\st j}}{M^2}\left(
\widetilde\Phi_{\{\partial G\}}^{\,ij}{-}\frac{1}{2}\,g_\st^{ij}\,\widetilde\Phi_{\{\partial{\cdot}G\}}
\right).
\label{e:doublepartialG}
\ee 
In Eq.~\ref{e:doublepartialG} we introduced the notation 
$\partial{\cdot}G$ in the subscript of one of the correlators to indicate 
that these two operators in this correlator have been contracted. Also for double 
$p_\st$-weighting we write down (selected contributions in)
the unintegrated result starting with Eq.~\ref{TMDstructure}.
We find for the double weighted result,
\bea
\frac{p_\st^\alpha p_\st^\beta}{M^2}\Phi^{[U]}(x,p_\st) &\ =\ &
\frac{p_{\st}^\alpha p_{\st}^\beta}{M^2}\,\Phi(x,p_\st^2)
-\frac{1}{2}g_\st^{\alpha\beta}
\,\frac{p_{\st i}}{M}\widetilde \Phi_{\partial}^{i (1)}(x,p_\st^2)
-\frac{1}{2}\pi C_G^{[U]}g_\st^{\alpha\beta}
\,\frac{p_{\st i}}{M} \Phi_{G}^{i (1)}(x,p_\st^2)
\nonumber \\ &\quad -&
\frac{1}{2M}
\,p_{\st}^{\{\alpha}
\,\widetilde\Phi_\partial^{\beta\} (1)}(x,p_\st^2)
-\pi C_G^{[U]}\frac{1}{2M}
\,p_{\st}^{\{\alpha}\,\Phi_G^{\beta\} (1)}(x,p_\st^2)
+ \ldots
\nonumber \\[3pt] &\quad +&
\widetilde\Phi_{\partial\partial}^{\alpha\beta (2)}(x,p_\st^2)
+\pi C_G^{[U]}
\,\widetilde\Phi_{\{\partial G\}}^{\alpha\beta (2)}(x,p_\st^2)
+\pi^2 C_{GG,c}^{[U]}
\,\Phi_{GG,c}^{\alpha\beta (2)}(x,p_\st^2)
+ \ldots
\nonumber \\[3pt] &\quad -& \frac{1}{2}\,g_{\st}^{\alpha\beta}\left(
\widetilde\Phi_{\partial{\cdot}\partial}^{(2)}(x,p_\st^2) +
\pi C_G^{[U]}\,\widetilde\Phi_{\{\partial{\cdot}G\}}^{(2)}(x,p_\st^2) +
\pi^2 C_{GG,c}^{[U]}\,\Phi_{G{\cdot}G,c}^{(2)}(x,p_\st^2)\right)+\ldots \, ,
\label{fullsecond-1}
\eea
an equation that after symmetrizing in $\alpha$ and $\beta$ 
can be used to identify the remaining TMDs for a spin 1/2 target.
We have omitted terms with rank 1 tensors in $p_\st$
multiplying 
$\widetilde\Phi_{\partial\partial\partial}^{\alpha\beta i}$, etc., 
as well as terms with rank 3 or rank 4 tensors like 
$p_{\st\,\,\,i}^{\alpha\beta}$ and $p_{\st\,\,\,ij}^{\alpha\beta}$.
The last line in Eq.~\ref{fullsecond-1}, containing terms that arise by contraction of the indices $i$ and $j$ in the explicit rewriting of the product $p_\st^\alpha p_\st^\beta p_{\st ij}$ through the contribution $g_\st^{\alpha\beta}g_{\st ij}$, is included for completeness.
Note that taking the trace $-\frac{1}{2}g_{\st\alpha\beta}$ gives
\bea
-\frac{p_{\st}^2}{2M^2}\,\Phi^{[U]}(x,p_\st) &\ =\ &
\Phi^{(1)}(x,p_\st^2)
+\frac{p_{\st i}}{M}
\,\widetilde\Phi_\partial^{i (1)}(x,p_\st^2)
+\pi C_G^{[U]}
\,\frac{p_{\st i}}{M}\,\Phi_G^{i (1)}(x,p_\st^2)
\nonumber \\[3pt] &\quad -&
\frac{1}{2}\widetilde\Phi_{\partial{\cdot}\partial}^{(2)}(x,p_\st^2)
-\frac{1}{2} \pi C_G^{[U]}
\,\widetilde\Phi_{\{\partial{\cdot}G\}}^{(2)}(x,p_\st^2)
-\frac{1}{2}\pi^2 C_{GG,c}^{[U]}
\,\Phi_{G{\cdot}G,c}^{(2)}(x,p_\st^2)+\ldots
\nonumber \\[3pt] &\quad +&
\left(\widetilde\Phi_{\partial{\cdot}\partial}^{(2)}(x,p_\st^2)
+\pi C_G^{[U]}
\,\widetilde\Phi_{\{\partial{\cdot}G\}}^{(2)}(x,p_\st^2)
+\pi^2 C_{GG,c}^{[U]}
\,\Phi_{G{\cdot}G,c}^{(2)}(x,p_\st^2)
\right)+\ldots \, ,
\label{fullsecond-trace}
\eea
where the terms between brackets come from the correlators with contracted operators in Eq.~\ref{fullsecond-1}. This shows that
\bea
\frac{p_\st^{\alpha\beta}}{M^2}\,\Phi^{[U]}(x,p_\st) &\ =\ &
\frac{p_{\st}^{\alpha\beta}}{M^2}\,\Phi(x,p_\st^2)
\nonumber \\ &\quad -&
\frac{1}{2M}
\left(p_{\st}^{\{\alpha}
\,\widetilde\Phi_\partial^{\beta\} (1)}(x,p_\st^2) 
- {\rm trace}\right)
-\frac{1}{2M}\,\pi C_G^{[U]}
\left(\,p_{\st}^{\{\alpha}
\,\Phi_G^{\beta\} (1)}(x,p_\st^2)
- {\rm trace}\right)
+ \ldots
\nonumber \\[3pt] &\quad +&
\left(\widetilde\Phi_{\partial\partial}^{\alpha\beta (2)}(x,p_\st^2)
- {\rm traces}\right)
+\pi C_G^{[U]}
\left(\widetilde\Phi_{\{\partial G\}}^{\alpha\beta (2)}(x,p_\st^2)
- {\rm traces}\right)
\nonumber \\ &\quad +&
\pi^2 C_{GG,c}^{[U]}
\left(\Phi_{GG,c}^{\alpha\beta (2)}(x,p_\st^2)
- {\rm traces}_c\right)
+\ldots \, ,
\label{fullsecond-2}
\eea
illustrating how the projection with the properly symmetrized
traceless second rank tensor $p_\st^{\alpha\beta}$ gives the properly 
symmetrized traceless TMD structures.
The terms without azimuthal dependence are identified with
the {\em rank two} TMD correlators, which for a spin 1/2 target
are parametrized as
\bea
&&
\frac{p_{\st ij}}{M^2}\widetilde\Phi_{\partial\partial}^{ij}(x,p_\st^2)
= h_{1T}^{\perp (A)}(x,p_\st^2)
\,\frac{p_{\st ij}S_\st^i\,\gamma_5\gamma_{\st}^j}{M^2}
\,\frac{\slashed{P}}{2},
\\ &&
\frac{p_{\st ij}}{M^2}\Phi_{GG,1}^{ij}(x,p_\st^2)
= \frac{1}{\pi^2}h_{1T}^{\perp (B1)}(x,p_\st^2)
\,\frac{p_{\st ij}S_\st^i\,\gamma_5\gamma_{\st}^j}{M^2}
\,\frac{\slashed{P}}{2},
\\ &&
\frac{p_{\st ij}}{M^2}\Phi_{GG,2}^{ij}(x,p_\st^2)
= \frac{1}{\pi^2}h_{1T}^{\perp (B2)}(x,p_\st^2)
\,\frac{p_{\st ij}S_\st^i\,\gamma_5\gamma_{\st}^j}{M^2}
\,\frac{\slashed{P}}{2},
\\ &&
\frac{p_{\st ij}}{M^2}\widetilde\Phi_{\{\partial G\}}^{ij}(x,p_\st^2)
= 0.
\eea
The last TMD correlator in these equations is a T-odd rank 2 TMD 
correlator, which is not present for a spin 1/2 target.
The result can also be summarized as the existence of {\em three}
universal pretzelocity functions $h_{1T}^{\perp (A)}$, $h_{1T}^{\perp (B1)}$ and
$h_{1T}^{\perp (B2)}$ and a gauge link dependence given by
\be 
h_{1T}^{\perp [U]}(x,p_\st^2)
= h_{1T}^{\perp (A)}(x,p_\st^2)
+ C_{GG,1}^{[U]}\,h_{1T}^{\perp (B1)}(x,p_\st^2)
+ C_{GG,2}^{[U]}\,h_{1T}^{\perp (B2)}(x,p_\st^2).
\ee
This shows e.g.\ that $h_{1T}^{\perp [{\rm SIDIS}]}(x,p_\st^2)
=h_{1T}^{\perp [{\rm DY}]}(x,p_\st^2)$, but that for other
processes (with more complicated gauge links) other
combinations of the three possible pretzelocity functions occur.
In asymmetries involving $p_\st^{\alpha\beta}$-moments of the quark TMD 
correlator contributions from all four correlators can appear.
In particular we find for a transversely polarized spin 1/2 target three pretzelocity functions, 
as was already established in Eq.~\ref{threepretzel}. 
For a spin 1/2 target our treatment is complete, since there are no
higher rank TMD correlators such as 
$p_{\st ijk}\widetilde\Phi_{\partial\partial\partial}^{\ ijk}
(x,p_\st^2)$. In the case of a spin 1/2 target the pretzelocity TMD functions $h_{1T}^{\perp (B2)}$ 
actually was referred to as junk TMD
in Ref.~\cite{Bomhof:2007xt}.

We want to summarize our results in this section in tabular
form. We first represent the contributions in Eq.~\ref{TMDstructure}
in Table~\ref{t:spinhalfcol}. The assignment of the TMD PDFs for
an unpolarized and polarized spin 1/2 target has been discussed in
this section and is summarized in Tables~\ref{t:unpol} 
and~\ref{t:spinhalfpol}.
For the corresponding fragmentation functions the assignments are different, 
since gluonic pole
matrix elements vanish in that case~\cite{Gamberg:2010gp,Meissner:2008yf,Gamberg:2008yt} and all functions are
assigned to $\widetilde\Phi_{\partial\ldots\partial}$ operator
structures. The assignments thus are as in 
Tables~\ref{t:spinhalffragunp} and~\ref{t:spinhalffragpol}. 
The T-odd TMD PFFs (such as the Collins function $H_1^\perp$) 
are due to the fact that the definitions of fragmentation
functions involve non-plane wave states or equivalently
a hadronic number operator, which are not invariant 
under time-reversal. 
There, thus, is only a single (T-even) function 
$H_{1T}^\perp(z,k_\st^2)$ appearing in the parametrization of the 
correlator $\Delta_{\partial\partial}^{\alpha\beta}(x,p_\st^2)$.

\begin{table}[!tb]
\centering
\begin{tabular}{|m{26mm}|m{26mm}|m{26mm}|m{26mm}|m{26mm}|}
\hline
\multicolumn{4}{|c|}{GLUONIC POLE RANK} \\ \hline
\qquad\quad 0 & \qquad\quad 1 & \qquad\quad 2 & \qquad\quad 3 
\\ \hline
$\Phi(x,p_\st^2)$
&$\pi C_{G}^{[U]}\,\Phi_{G}$
&$\pi^2 C_{GG,c}^{[U]}\,\Phi_{GG,c}$
&$\pi^3 C_{GGG,c}^{[U]}\,\Phi_{GGG,c}$
\\[2pt]
\hline
$\widetilde\Phi_\partial$
&$\pi C_{G}^{[U]}\,\widetilde\Phi_{\{\partial G\}}$
&$\pi^2 C_{GG,c}^{[U]}\,\widetilde\Phi_{\{\partial GG\},c}$
& \ldots
\\[2pt]
\hline
$\widetilde\Phi_{\partial\partial}$
&$\pi C_{G}^{[U]}\,\widetilde\Phi_{\{\partial\partial G\}}$
& \ldots & \ldots
\\[2pt]
\hline
$\widetilde\Phi_{\partial\partial\partial}$
& \ldots & \ldots & \ldots
\\[2pt] 
\hline
\end{tabular}
\parbox{0.85\textwidth}{
\caption{
The contributions in the TMD correlator for correlators ordered 
in columns according to the number of gluonic poles ($G$)
and ordered in rows according to the number of contributing partial derivatives ($\partial$ = $D-A$). The rank of these operators is
equal to the sum of these numbers. Their twist is equal to the
rank + 2.
\label{t:spinhalfcol}}}
\end{table}

\begin{table}[!tb]
\begin{minipage}{0.45\linewidth}
\centering
\begin{tabular}{|m{15mm}|m{15mm}|m{15mm}|}
\hline
\multicolumn{3}{|c|}{PDFs FOR SPIN 0 HADRONS} \\ \hline
$f_1$
&$h_{1}^{\perp}$
&
\\[2pt]
\hline
&
\\[2pt]
\cline{1-2}
\\[2pt]
\cline{1-1}
\end{tabular}
\parbox{0.85\textwidth}{
\caption{
The assignment of TMD PDFs for a spin 0 or unpolarized target 
to the quark 
correlators as given in Table~\ref{t:spinhalfcol} involve at most 
rank 1 TMD correlators. There is no T-even function corresponding
to $\widetilde\Phi_\partial^i$.
\\ {}
\label{t:unpol}}}
\end{minipage}
\begin{minipage}{0.45\linewidth}
\centering
\begin{tabular}{|m{13mm}|m{12mm}|m{21mm}|}
\hline
\multicolumn{3}{|c|}{PDFs FOR SPIN 1/2 HADRONS} \\ \hline
$g_{1}$, $h_{1}$
&$f_{1T}^{\perp}$
&$h_{1T}^{\perp(B1)}$, $h_{1T}^{\perp(B2)}$
\\[2pt]
\hline
$g_{1T}$, $h_{1L}^{\perp}$
&
\\[2pt]
\cline{1-2}
$h_{1T}^{\perp(A)}$
\\[1pt]
\cline{1-1}
\end{tabular}
\parbox{0.85\textwidth}{
\caption{
The assignment of TMD PDFs for a polarized spin 1/2 target to the quark 
correlators as given in Table~\ref{t:spinhalfcol} involve at most 
rank 1 TMD correlators for longitudinal polarization, while
they involve also rank 2 TMD correlators for a transversely
polarized spin 1/2 target.
\label{t:spinhalfpol}}}
\end{minipage}
\\[6mm]
\begin{minipage}{0.45\linewidth}
\centering
\begin{tabular}{|m{15mm}|m{15mm}|m{15mm}|}
\hline
\multicolumn{3}{|c|}{PFFs FOR SPIN 0 HADRONS} \\ \hline
$D_1$
&
&
\\[2pt]
\hline
$H_{1}^{\perp}$
&
\\[2pt]
\cline{1-2}
\\[1pt]
\cline{1-1}
\end{tabular}
\parbox{0.85\textwidth}{
\caption{
The operator structure of quark TMD PFFs for spin 0 or unpolarized
hadrons. All gluonic pole matrix elements vanish. 
\label{t:spinhalffragunp}}}
\end{minipage}
\begin{minipage}{0.45\linewidth}
\centering
\begin{tabular}{|m{22mm}|m{11.5mm}|m{11.5mm}|}
\hline
\multicolumn{3}{|c|}{PFFs FOR SPIN 1/2 HADRONS} \\ \hline
$G_{1}$, $H_{1}$
&
&
\\[2pt]
\hline
$G_{1T}$, $H_{1L}^{\perp}$, $D_{1T}^{\perp}$
&
\\[2pt]
\cline{1-2}
$H_{1T}^{\perp}$
\\[1pt]
\cline{1-1}
\end{tabular}
\parbox{0.85\textwidth}{
\caption{
The operator structure of quark TMD PFFs for polarized
spin 1/2 hadrons. Gluonic pole matrix elements vanish.
\label{t:spinhalffragpol}}}
\end{minipage}
\end{table}

\subsection{Results for spin 1 hadrons}

Extension to higher spin targets is straightforward. 
We illustrate this by giving in Table~\ref{t:spinone} 
the assignments for spin 1 tensor polarized TMD functions. These were first given in Ref.~\cite{Bacchetta:2000jk}. The (slightly updated) parametrization of the TMD correlator for the TMD PDFs for a tensor polarized target are given in Appendix~\ref{A:spinonedis} as well as the parametrization of the TMD PFFs in Appendix~\ref{A:spinonefrag}.
From these tensor polarized spin 1 contributions, the $f_{1TT}^{[U]}(x,p_\st^2)$ and $h_{1TT}^{\perp [U]}(x,p_\st^2)$ can be written as a combination of multiple universal PDFs, multiplied with process dependent gluonic pole factors,
\bea
f_{1TT}^{[U]}(x,p_\st^2)
&=& f_{1TT}^{(A)}(x,p_\st^2)
+ C_{GG,c}^{[U]}\,f_{1TT}^{(Bc)}(x,p_\st^2), \\
h_{1TT}^{\perp [U]}(x,p_\st^2)
&=& C_{G}^{[U]}\,h_{1TT}^{\perp (A)}(x,p_\st^2)
+ C_{GGG,c}^{[U]}\,h_{1TT}^{\perp (Bc)}(x,p_\st^2).
\eea
Note that the $h_{1TT}^{\perp [U]}(x,p_\st^2)$ is a rank 3 object, for 
which all contributing universal functions are multiplied with a 
process dependent prefactor. A special case is the T-odd TMD PDF 
$h_{1LT}$, which is forbidden because of time-reversal invariance. 
Following Ref.~\cite{Bacchetta:2000jk}, this rank 0 TMD PDF is 
defined as the combination
$h_{1LT}(x,p_\st^2) = h_{1LT}^{\prime}(x,p_\st^2)
+h_{1LT}^{\perp (1)}(x,p_\st^2)$ and is shown as the wiped-out
function in Table~\ref{t:spinone}. It shows a nice feature
of our TMD functions of definite rank. In the first column
only T-even TMD PDFs are allowed, in the second column only T-odd
ones, etc. The first victim of the application of 
time-reversal invariance for leading quark TMDs, thus, is 
$h_{1LT}(x,p_\st^2)$, a (T-forbidden) transversely polarized quark 
distribution function in a tensor polarized hadron.
Note that the rank 2, T-odd function $h_{1LT}^{\perp}(x,p_\st^2)$ 
is allowed. The only rank 0 function for
a tensor polarized spin 1 target thus is $f_{1LL}(x,p_\st^2)$, introduced
as the distribution $b_1$ in Ref.~\cite{Hoodbhoy:1988am}.

For fragmentation functions, gluonic pole contributions all
vanish and only the first column survives. The parametrization
of the higher rank correlators contain the T-even and T-odd
TMD fragmentation functions. The fragmentation functions describing
fragmentation into a tensor polarized hadron are given in
Table~\ref{t:spinonefrag}.

\begin{table}[t]
\centering
\begin{tabular}{|p{25mm}|p{25mm}|p{25mm}|p{25mm}|}
\hline
\multicolumn{4}{|c|}{PDFs FOR TENSOR POLARIZED SPIN 1 HADRONS} \\ 
\hline
$f_{1LL}$, $\xcancel{h_{1LT}}$
&$h_{1LL}^{\perp}$, $g_{1LT}$, $h_{1TT}$
&$f_{1TT}^{(Bc)}$
&$h_{1TT}^{\perp(Bc)}$
\\[2pt]
\hline
$f_{1LT}$
&$h_{1LT}^{\perp}$, $g_{1TT}$
&
\\[2pt]
\cline{1-3}
$f_{1TT}^{(A)}$
&$h_{1TT}^{\perp(A)}$
\\[2pt]
\cline{1-2}
\\[1pt]
\cline{1-1}
\end{tabular}
\parbox{0.85\textwidth}{
\caption{
The operator assignments of TMD PDFs for a tensor polarized spin 
1 target require operator structures up to rank 3. 
There are several different functions $f_{1TT}(x,p_\st^2)$ and
$h_{1TT}^\perp(x,p_\st^2)$. 
\label{t:spinone}}}
\end{table}

 \begin{table}[!tb]
\centering
\begin{tabular}{|p{37mm}|p{21mm}|p{21mm}|p{21mm}|}
\hline
\multicolumn{4}{|c|}{PFFs FOR TENSOR POLARIZED SPIN 1 HADRONS} \\ 
\hline
$D_{1LL}$, $H_{1LT}$
&
&
&
\\[2pt]
\hline
$D_{1LT}$, $H_{1LL}^{\perp}$, $G_{1LT}$, $H_{1TT}$
&
&
\\[2pt]
\cline{1-3}
$D_{1TT}$, $H_{1LT}^{\perp}$, $G_{1TT}$
&
\\[2pt]
\cline{1-2}
$H_{1TT}^{\perp}$
\\[1pt]
\cline{1-1}
\end{tabular}
\parbox{0.85\textwidth}{
\caption{
The operator structure of TMD PFFs for a tensor polarized spin 
1 target requires operator structures up to rank 3.
\label{t:spinonefrag}}}
\end{table}

\subsection{Bessel weights}

We note that the TMDs $f^{(m)}(x,p_{\st}^{2})$ of a given rank do not contain operators of
definite twist. This is only true for transverse moments 
$f^{(m)}_{\ldots} (x)$ after $p_\st$-integration. The TMD correlators
of definite rank appearing in the parametrization 
in Eq.~\ref{TMDstructure} only are integrated over azimuthal
directions. The rank just refers to the azimuthal dependence
of the correlators in the full correlator $\Phi^{[U]}(x,p_\st)$.

Using that for a given rank $m$, there are
two independent combinations $p_\st^{i_1\ldots i_m}$
$\propto$ $\vert p_\st\vert^m\,\exp(\pm im\varphi_p)$, 
it is equivalent to consider
\be
\frac{p_{\st i_1\ldots i_m}}{M^m}
\,\widetilde\Phi_{\ldots}^{i_1\ldots i_m}(x, p_\st^2)
\qquad \mbox{or} \qquad
\widetilde \Phi_{\ldots}^{(m/2)}(x,p_\st^2)
\,e^{im\varphi_p},
\label{comparison}
\ee
where $\widetilde \Phi_{\ldots}^{(m/2)}(x,p_\st^2)$ = 
$(-p_\st^2/2M^2)^{m/2}\,\widetilde\Phi_{\ldots}(x,p_\st^2)$ 
assures the appropriate small $p_\st$-behavior. A 
suitable normalization of the correlator has
to assure that $\widetilde\Phi^{(m)}_{\ldots}(x,p_\st^2)$
reproduces the collinear transverse moments upon integration,
\be 
\widetilde\Phi_{\ldots}^{(m)}(x) 
= \int_0^\infty 2\pi\vert p_\st\vert\,d\vert p_\st\vert
\ \widetilde\Phi^{(m)}_{\ldots}(x,p_\st^2).
\ee
Knowing the correlators in Eq.~\ref{comparison} 
to be Fourier transforms of
nonlocal matrix elements in transverse space, 
it is natural to write the appropriately weighted TMD
PDF in their parametrization as a Bessel transform, 
\be 
\widetilde f^{(m/2)}_{\ldots}(x,\vert p_\st\vert)
= \int_0^\infty db\ \sqrt{\vert p_\st\vert b}
\,J_m(\vert p_\st\vert b)\,f^{(m/2)}_{\ldots}(x,b),
\ee
such that $f_{\ldots}^{(m/2)}(x,b)\,\exp(im\varphi_b)$ is 
the (two-dimensional) Fourier transform of 
$\widetilde f_{\ldots}^{(m/2)}(x,\vert p_\st\vert)
\,\exp(im\varphi_p)$. Bessel weightings are extensively studied in Ref.~\cite{Boer:2011xd}.

Bessel weighting may also offer a convenient way to incorporate the soft factor which usually is given in b-space~\cite{Collins:2011zzd}. This factor has been omitted from Eq.~\ref{e:qcor_O1O2}. Our decomposition in Eq.~\ref{TMDstructure}, however, can always be written down, but the $\Phi_{\ldots}(x,p_\st^2)$ will be modified by the inclusion of the soft factor.

\section{Conclusions}
In Eq.~\ref{TMDstructure} we have presented a parametrization 
for TMD quark correlators that
distinguishes different azimuthal dependences. For this we write down an
expansion in terms of irreducible tensors in the transverse 
momentum multiplied with correlators depending on $x$ and $p_\st^2$.
These correlators contain tensors describing the
polarization of the target and TMD functions depending on $x$ and
$p_\st^2$. The rank of the irreducible tensors in transverse momentum
space also defines the rank of the correlators and TMD functions 
multiplying this tensor. The field theoretical expression for the 
quark correlator basically has two quark fields connected by a 
gauge link. The operator structure of the TMD correlators of a 
definite rank contain Dirac gamma matrices, derivatives or 
gauge fields with transverse indices in color gauge-invariant
combinations and a definite rank. They are structured in a similar way 
as higher twist operators in the collinear case. Each independent operator
combination defines a particular TMD function. For leading
twist operators the relevant `transverse' operators are either
gluonic pole $(G)$ or partial derivative $(\partial=D-A)$
or combinations thereof. A special feature of these operator
combinations of definite rank is that they are either T-even 
or T-odd after extraction of a gluonic pole factor without
having to perform weighting and integration over transverse momentum.

Using the parametrization of TMD correlators in Eq.~\ref{TMDstructure}, 
one finds that rank 0 and rank 1 contributions are similar to
previously used definitions for T-even and T-odd contributions, 
such as e.g.\ obtained by combining `opposite' gauge links in Eq.~\ref{e:DEFINITE}. Just
as the collinear transverse moments, these TMD functions are
universal functions, multiplied with a process dependent prefactor, 
rather than nonuniversal gauge link dependent functions.
By explicitly and systematically looking at all contributing functions, 
one finds for an unpolarized target two TMD quark correlators,
the first being of rank 0, containing the TMD PDF $f_1(x,p_\st^2)$. 
Looking at the operator structure, it is interpreted as the momentum
distribution of quarks. The second unpolarized TMD correlator is a 
T-odd gluonic pole matrix element of rank 1. 
It contains the Boer-Mulders TMD PDF $h_1^\perp(x,p_\st^2)$.

For a polarized spin 1/2 target there are an additional eight
TMD correlators, containing rank 0, rank 1 and rank 2 contributions.
The two rank 0 correlators contain the TMD PDFs that are interpreted
as the well-known polarized spin distribution functions in 
longitudinally or transversely polarized targets. 
For a longitudinally polarized target, there is a T-even rank 1
TMD correlator $\widetilde\Phi_{\partial}(x,p_\st^2)$
containing the worm gear function $h_{1L}^\perp(x,p_\st^2)$.
For a transversely polarized target, there exist one T-even
TMD correlator $\widetilde\Phi_{\partial}(x,p_\st^2)$ and a T-odd
rank 1 TMD correlator $\Phi_{G}(x,p_\st^2)$ containing
the worm gear function $g_{1T}(x,p_\st^2)$ and the Sivers function
$f_{1T}^{\perp}(x,p_\st^2)$, respectively.
The three TMD correlators of rank 2 appear in the T-even correlator 
$\widetilde\Phi_{\partial\partial}(x,p_\st^2)$ and the two T-even double gluonic pole correlators
$\Phi_{GG,c}(x,p_\st^2)$, giving rise to the pretzelocity functions
$h_{1T}^{\perp (A)}(x,p_\st^2)$, $h_{1T}^{\perp (B1)}(x,p_\st^2)$ and
$h_{1T}^{\perp (B2)}(x,p_\st^2)$. These functions in general both 
show up in particular azimuthal asymmetries but with 
gauge link dependent prefactors, where the gauge link in turn
depends on the process. As for the functions themselves and in
particular their interpretation, the function
$h_{1T}^{\perp (A)}$ is related to the quark structure 
of a nucleon, while the functions $h_{1T}^{\perp (Bc)}$ 
are the ones involving quark-gluon correlations.

For a spin 1 target one finds apart from the above mentioned TMD 
correlators additional correlators because one also has the
possibility of tensor polarization. The full list of TMDs has
been given in Table~\ref{t:spinone}, including rank 3 
contributions coming with process dependent prefactors. Rank 3 
contributions like this are specific for targets with 
spin 1 or higher.

The procedure for defining universal TMD correlators of definite 
rank can be extended to gluon TMDs and to higher twist situations. 
The extension to gluon TMDs will be presented in a forthcoming
publication. The situation for higher twist TMDs
is complicated by the fact that the lowest twist operators that 
contribute to the TMDs not only contain two quark fields or 
two gluon fields, but they also contain additional (gluon) 
operators with transverse directions, $D_\st^\alpha$ and 
$F_\st^{n\alpha}$, no longer in the combination 
$\partial_\st^\alpha$.

As a final advantage of the universal TMD correlators we
mention that, although their nonlocal operator structure is of the 
form in Eq.~\ref{e:qcor_O1O2} with a particular gauge link $U$,
the TMD correlators of definite rank have definite T-behavior
(even or odd) and are independent of the gauge link. The $U$-dependence is in the gluonic pole factors and the color
structure of the operator combination. Thus one can study
the universal TMD correlators, for instance in lattice calculations, by
using just the sum and difference of the simplest
$U^{[+]}$ and $U^{[-]}$ staple links.

\section*{Acknowledgements}
This research is part of the research program of the ``Stichting 
voor Fundamenteel Onderzoek der Materie (FOM)", which is 
financially supported by the ``Nederlandse Organisatie voor
Wetenschappelijk Onderzoek (NWO)". MGAB also acknowledges
support of the FP7 EU-programme HadronPhysics3 (contract
no 283286). AM thanks the Alexander von Humboldt
Fellowship for Experienced Researchers, Germany, for support. 
AM also acknowledges support from the FOM
Programme ``Theoretical Physics in the LHC Era", for
visiting VU University and Nikhef, Amsterdam, where this work was initiated. 
We would like to acknowledge discussions with several participants 
of the QCD Evolution Workshop at Jefferson Lab (14-17 May 2012).
All figures were made using JaxoDraw~\cite{Binosi:2003yf,Binosi:2008ig}.

\appendix
\section{\label{A:matrix}Double weighted collinear matrix elements}
Double $p_\st$-weighting is worked out in the same way as
the single weighting. In evaluating the
weighting one can actually choose the derivatives to work on
$\xi$ or $0$ or mix these, but that does not matter for the
final answer. Since we just are interested in the general
structure, we work with the multi-parton element with all
gluons on the left side of the cut. For the gauge link
$U^{[+]}_{[0,\xi]}$ we note that (writing only the relevant
pieces of the matrix elements)~\cite{Buffing:2011mj}
\be 
i\partial_\st^\alpha U^{[+]}_{[0,\xi]}
= U^{[n]}_{[0,\infty]}\,i\partial_\st^\alpha U^\st_{[0_\st,\xi_\st]}
\,U^{[n]}_{[\infty,\xi]}
= U^{[n]}_{[0,\infty]}\,U^\st_{[0_\st,\xi_\st]}
\,iD_\st^\alpha\,U^{[n]}_{[\infty,\xi]},
\ee
which is then further evaluated using
\bea
iD_{\st}^{\alpha}\,U_{[\infty,\xi]}^{[n]}\ldots \psi(\xi)=U_{[\infty,\xi]}^{[n]}\bigg(iD_{\st}^{\alpha}(\xi)-A_{\st}^{\alpha}(\xi)+\pi \widetilde G^{n\alpha}(\xi)\bigg)\ldots \psi(\xi).
\label{e:TM2}
\eea
The $A_{\st}^{\alpha}(\xi)$ and $\pi \widetilde G^{n\alpha}(\xi)$ are defined as
\bea
&&A_{\st}^{\alpha}(\xi)=\frac{1}{2}\int_{-\infty}^{\infty}
d\eta{\cdot}P\ \epsilon(\xi{\cdot}P-\eta{\cdot}P)
\,U_{[\xi,\eta]}^{[n]} G^{n\alpha}(\eta)U_{[\eta,\xi]}^{[n]}, \label{e:defA} \\
&&\pi \widetilde G^{n\alpha}(\xi)=\frac{1}{2}\int_{-\infty}^{\infty}
d\eta{\cdot}P\ U_{[\xi,\eta]}^{[n]}G^{n\alpha}(\eta)
U_{[\eta,\xi]}^{[n]},
\label{e:defG}
\eea
with $\epsilon (\zeta)$ being the sign function taking the 
values $1$, $-1$ and $0$. Note that $\widetilde G^{n\alpha}(\xi)$
= $\widetilde G^{n\alpha}(\xi{\cdot}P,\xi_\st)$ does not depend
on $\xi{\cdot}n$. The above described method for calculating 
the structure of the matrix elements for a single transverse weighting 
can be extended to higher transverse weightings by repeated application 
of Eq.~\ref{e:TM2}. For example, for the $U^{[+]}$ gauge link this 
implies
\bea
&&iD_{\st}^{\alpha}\,iD_{\st}^{\beta}\,U_{[\infty,\xi]}^{[n]}\ldots 
\psi(\xi)
\nonumber \\ &&\mbox{}\hspace{1cm}\mbox{}
= U_{[\infty,\xi]}^{[n]} 
\bigg(\Big(iD_{\st}^{\alpha}(\xi)-A_{\st}^{\alpha}(\xi)\Big)
\Big(iD_{\st}^{\beta}(\xi)-A_{\st}^{\beta}(\xi)\Big)\bigg)\ldots 
\psi(\xi)
+U_{[\infty,\xi]}^{[n]}\bigg(\pi \widetilde G^{n\alpha}(\xi)
\,\pi \widetilde G^{n\beta}(\xi)\bigg)\ldots \psi(\xi) 
\nonumber \\&&\mbox{}\hspace{1.5cm}\mbox{}
+U_{[\infty,\xi]}^{[n]}\bigg(\pi \widetilde G^{n\alpha}(\xi) 
\Big(iD_{\st}^{\beta}(\xi)-A_{\st}^{\beta}(\xi)\Big)\bigg)\ldots 
\psi(\xi)
+U_{[\infty,\xi]}^{[n]}\bigg(
\Big(iD_{\st}^{\alpha}(\xi)-A_{\st}^{\alpha}(\xi)\Big)\pi \widetilde G^{n\beta}(\xi)\bigg)\ldots \psi(\xi). \nonumber \\ &&
\label{e:doubleD_1}
\eea
After integration over $p_\st$ the resulting correlators can
be rewritten in terms of color gauge-invariant multi-parton
correlators as was done for the single weighting. In this case
one needs the correlators of the form
\bea
&&
\Phi_{O_1O_2}^{\alpha\beta}(x-x_{1}-x_{2},x_{1},x_{2}|x)=\int\frac{d\,\xi{\cdot}P}{2\pi}\frac{d\,\eta{\cdot}P}{2\pi}\frac{d\,\eta^{\prime}{\cdot}P}{2\pi}e^{i x_{2}(\eta^{\prime}\cdot P)}e^{i x_{1}(\eta\cdot P)}e^{i(x-x_{1}-x_{2})(\xi\cdot P)}
\nonumber \\
&&\hspace{48mm} \times \langle P,S|\overline{\psi}(0)U^{[n]}_{[0,\eta^{\prime}]}O_{1\,\st}^{\alpha}(\eta^{\prime})U^{[n]}_{[\eta^{\prime},\eta]}O_{2\,\st}^{\beta}(\eta)U^{[n]}_{[\eta,\xi]}\psi(\xi)|P,S\rangle\rfloor_{LC} \label{e:qcor_AB},
\eea
with $O_{1\,\st}^\alpha$ and $O_{2\,\st}^\alpha$ hermitian 
operators like 
$iD_\st^\alpha$ and/or $F_\st^{n\alpha}$. 
In general more than one color structure is possible. As an example, for the operator combination $\overline\psi FF\psi$, one can have the two distinct color configurations
\bea
c=1:\hspace{4mm}&&\tr_c\left[FF\psi\overline\psi\,\right]=\overline\psi FF\psi =\overline\psi{}^r F^{rs^{\prime}}F^{s^{\prime}s}\psi^s , \\
c=2:\hspace{4mm}&&\tr_c\left[FF\right]\tr_c\left[\psi\overline\psi\,\right]=\overline\psi \psi\tr_c\left[FF\right]=\overline\psi{}^r \psi^r F^{ss^{\prime}}F^{s^{\prime}s}.
\eea
An example of a correlator one needs is 
\begin{eqnarray}
\Phi_{DD}^{\alpha\beta}(x)
&=&
\int\frac{d\,\xi{\cdot}P}{2\pi}e^{i x(\xi \cdot P)}\langle P,S|\overline{\psi}(0)U^{[n]}_{[0,\xi]}iD_{\st}^{\alpha }(\xi)iD_{\st}^{\beta}(\xi)\psi(\xi)|P,S\rangle\rfloor_{LC} \label{e:qcor_DD} \nonumber \\
&=& \int dx_1\,dx_2\ \Phi_{DD}^{\alpha\beta}(x-x_1-x_2,x_1,x_2\vert x).
\eea
Other correlators involving $A_\st^\alpha(\xi)$ or 
$\pi\widetilde G_\st^{n\alpha}(\xi)$, given in Eqs~\ref{e:defA} and \ref{e:defG}, are given by 
\begin{eqnarray}
&&\Phi_{AA}^{\alpha\beta}(x)=\int dx_{1} \,\text{PV} 
\frac{i}{x_{1}}\int dx_{2} \,\text{PV} \frac{i}{x_{2}}\Phi_{FF}^{\alpha\beta}(x-x_{1}-x_{2},x_{1},x_{2}|x), 
\label{e:qcor_AA} 
\\
&&\Phi_{AD}^{\alpha\beta}(x)=\int dx_{1}dx_2 \,\text{PV} 
\frac{i}{x_{1}}\Phi_{FD}^{\alpha\beta}(x-x_{1}-x_{2},x_{1},x_{2}|x), 
\label{e:qcor_AD} 
\\
&&\Phi_{DA}^{\alpha\beta}(x)=\int dx_{1}dx_2 \,\text{PV} 
\frac{i}{x_{2}}\Phi_{DF}^{\alpha\beta}(x-x_{1}-x_{2},x_{1},x_{2}|x),
\label{e:qcor_DA}
\\
&&\Phi_{GD}^{\alpha\beta}(x)=\int dx_{2}\,
\Phi_{FD}^{\alpha\beta}(x-x_{2},0,x_{2}|x), 
\label{e:qcor_GD} 
\\
&&\Phi_{DG}^{\alpha\beta}(x)=\int dx_{1} \,
\Phi_{DF}^{\alpha\beta}(x-x_{1},x_{1},0|x), 
\label{e:qcor_DG} 
\\
&&\Phi_{GA}^{\alpha\beta}(x)=\int dx_2 \,\text{PV} 
\frac{i}{x_{2}}\Phi_{FF}^{\alpha\beta}(x-x_{2},0,x_{2}|x), 
\label{e:qcor_GA} 
\\
&&\Phi_{AG}^{\alpha\beta}(x)=\int dx_{1} \,\text{PV} 
\frac{i}{x_{1}}\Phi_{FF}^{\alpha\beta}(x-x_{1},x_{1},0|x), 
\label{e:qcor_AG} 
\\
&&\Phi_{GG,c}^{\alpha\beta}(x)
=\Phi_{FF,c}^{\alpha\beta}(x,0,0|x).
\label{e:qcor_GG}
\end{eqnarray}
Using for the correlators, just as for the single weighted case, the
notation $\widetilde\Phi_{\partial\ldots}$ for the correlators with 
covariant derivative minus a correlator with a principal value integration ($iD_\st - A_\st$), implies
\bea
\widetilde\Phi^{\alpha\beta}_{\partial\partial}(x)&=&\Phi^{\alpha\beta}_{DD}(x)
-\Phi^{\alpha\beta}_{DA}(x)
-\Phi^{\alpha\beta}_{AD}(x)
+\Phi^{\alpha\beta}_{AA}(x), \\
\widetilde\Phi_{\partial G}^{\alpha\beta}(x)
&=&\Phi_{DG}^{\alpha\beta}(x)-\Phi_{AG}^{\alpha\beta}(x), \\
\widetilde\Phi_{G\partial}^{\alpha\beta}(x)
&=&\Phi_{GD}^{\alpha\beta}(x)-\Phi_{GA}^{\alpha\beta}(x).
\eea
The second transverse moment in terms of the collinear functions
then is (symmetrizing in $\alpha$
and $\beta$)
\bea
\int d^2p_\st\ p_\st^{\{\alpha}p_\st^{\beta\}}\,\Phi^{[+]}(x,p_\st)
= \widetilde\Phi^{\left\{\alpha\beta\right\}}_{\partial\partial}(x)
+\pi\, \widetilde\Phi_{\partial G}^{\left\{\alpha\beta\right\}}(x)
+\pi\, \widetilde\Phi_{G\partial}^{\left\{\alpha\beta\right\}}(x)
+\pi^2 \, \Phi_{GG,1}^{\left\{\alpha\beta\right\}}(x),
\eea
which is the result given in Eq.~\ref{e:Phipp} with $C_{G}^{[+]}=1$, $C_{GG,1}^{[+]}=1$ 
and $C_{GG,2}^{[+]}=0$. For other gauge link structures, similar calculations can be performed.

\section{\label{A:matrixtransverse}Double weighted TMD matrix elements}

The leading relevant TMD operator structure for our
considerations referred to in Eq.~\ref{e:qcor_O1O2} is
bilocal,
\be 
\Phi^{[U]}(x,p_\st;n) 
= \left.\int \frac{d\xi{\cdot}P\,d^2\xi_\st}{(2\pi)^3}
\ e^{ip\cdot\xi}\,\langle P\vert \overline\psi(0)\,U_{[0,\xi]}
\,O (\xi)\vert P\rangle\right|_{\xi{\cdot}n=0}.
\ee
The nonlocality, however, involves a transverse separation,
hence the gauge link $U_{[0,\xi]}$ in general can be complicated.
For the two cases $c$ = 1, 2 one now finds for the
gauge link $U_{[0,\xi]}^{[(\Box)+]}$ the nonlocal structures
\bea
c=1:\hspace{4mm}&&\tr_c\left[U_{[0,\xi]}^{[(\Box)+]}
\widetilde G(\xi)\,\widetilde G(\xi)\psi(\xi)\overline\psi(0)\right]
=\overline\psi{}(0) U_{[0,\xi]}^{[(\Box)+]}
\widetilde G(\xi)
\widetilde G(\xi)
\psi(\xi) , \\
c=2:\hspace{4mm}&&\tr_c\left[U_{[0,\xi]}^{[\Box]}\widetilde G(\xi) 
\widetilde G(\xi) \right]\tr_c\left[U_{[0,\xi]}^{[+]}
\,\psi(\xi)\overline\psi(0)\,\right]
=\overline\psi{}(0)
\,U_{[0,\xi]}^{[+]}\,\psi(\xi)
\tr_c\left[U_{[0,\xi]}^{[+]}\widetilde G(\xi) 
\widetilde G(\xi) U_{[\xi,0]}^{[-]^{\dagger}} \right].
\eea

\section{\label{A:spinonedis}Parametrization of the spin 1 distribution correlator}
The parametrization of a distribution correlator for a spin 1 hadron 
was first given in Ref.~\cite{Bacchetta:2000jk} and is given by
\be
\Phi (x,p_{\st})=\Phi_{U}(x,p_{\st})+\Phi_{L}(x,p_{\st})
+\Phi_{T}(x,p_{\st})+\Phi_{LL}(x,p_{\st})
+\Phi_{LT}(x,p_{\st})+\Phi_{TT}(x,p_{\st}),
\ee
where the contributions $\Phi_{U}(x,p_{\st})$, $\Phi_{L}(x,p_{\st})$ 
and $\Phi_{T}(x,p_{\st})$ are parametrized in the same way as
those contributions in the correlators that describe an unpolarized, 
longitudinally or transversely polarized spin 1/2 particle, 
given in Eq.~\ref{e:par} using the notation of Ref.~\cite{Bacchetta:2006tn}. We update the 
parametrization for the remaining correlators contributing for spin 1 
particles, using the same TMD PDFs as in Ref.~\cite{Bacchetta:2000jk}. This leads for the leading twist TMDs to
\bea
\Phi_{LL}(x,p_{\st})&=&\bigg\{
f_{1LL}(x,p_\st^2)S_{\sL\sL}
+ih_{1LL}^{\perp}(x,p_\st^2)
\,S_{\sL\sL}\frac{\slashed{p}_{\st}}{M}
\bigg\}\frac{\slashed{P}}{2}, 
\\
\Phi_{LT}(x,p_{\st})&=&\bigg\{
-f_{1LT}(x,p_\st^2)\frac{p_{\st}\cdot S_{\sL\st}}{M}
+g_{1LT}(x,p_\st^2)\,\epsilon_{\st}^{\mu\nu}S_{\sL\st\,\mu}
\frac{p_{\st\,\nu}}{M}\gamma_{5} 
\nonumber \\
&&\hspace{5mm}
+h_{1LT}^{\prime}(x,p_\st^2)\gamma_{5}\gamma_{\nu}
\epsilon_{\st}^{\nu\rho}S_{\sL\st\,\rho}
-ih_{1LT}^{\perp}(x,p_\st^2)\frac{p_{\st}\cdot S_{\sL\st}}{M}
\frac{\slashed{p}_{\st}}{M}
\bigg\}\frac{\slashed{P}}{2}, 
\\
\Phi_{TT}(x,p_{\st})&=&\bigg\{
f_{1TT}(x,p_\st^2)
\frac{p_{\st\,\alpha\beta} S_{\st\st}^{\alpha\beta}}{M^2}
-g_{1TT}(x,p_\st^2)\epsilon_{\st}^{\mu\nu}
S_{\st\st\,\nu\rho}\frac{p_{\st}^{\rho}p_{\st\,\mu}}{M^2}\gamma_{5} 
\nonumber \\
&&\hspace{5mm}
-h_{1TT}^{\prime}(x,p_\st^2)\gamma_{5}\gamma_{\nu}
\epsilon_{\st}^{\nu\rho}S_{\st\st\,\rho\sigma}
\frac{p_{\st}^{\sigma}}{M}
+ih_{1TT}^{\perp}(x,p_\st^2)\frac{p_{\st\,\alpha\beta} S_{\st\st}^{\alpha\beta}}{M^2}\frac{\slashed{p}_{\st}}{M}
\bigg\}\frac{\slashed{P}}{2}.
\eea
We note that all polarized quark distributions ($g$ and $h$ functions)
in a tensor polarized target, $h_{1LL}^{\perp}$, $g_{1LT}$, $h_{1LT}^{\prime}$, $h_{1LT}^{\perp}$, $g_{1TT}$, $h_{1TT}^{\prime}$ and $h_{1TT}^{\perp}$, are T-odd.
Just as in Ref.~\cite{Bacchetta:2000jk}, the integrated case is given by
\bea
\Phi_{LL}(x)&=&
f_{1LL}(x)S_{\sL\sL}\frac{\slashed{P}}{2}, \\
\Phi_{LT}(x)&=&
h_{1LT}(x)\gamma_{\nu}
\epsilon_{\st}^{\nu\rho}S_{\sL\st\,\rho}\frac{\slashed{P}}{2}, 
\\
\Phi_{TT}(x)&=&0,
\eea
where
\bea
h_{1LT}(x,p_\st^2) &=& 
h_{1LT}^{\prime}(x,p_\st^2) + h_{1LT}^{\perp (1)}(x,p_\st^2). 
\eea
For identifying the proper rank 1 TMD, it is also useful to define
\bea
h_{1TT}(x,p_\st^2) &=& 
h_{1TT}^{\prime}(x,p_\st^2) + h_{1TT}^{\perp (1)}(x,p_\st^2).
\eea

\section{\label{A:spinonefrag}Parametrization of the fragmentation correlator}
The parametrization of the fragmentation correlator for a spin 1 
particle, also first used in Ref.~\cite{Bacchetta:2000jk}, 
is similar in structure to the parametrization of the distribution 
correlator and is given by
\be
\Delta (z,k_{\st})=\Delta_{U}(z,k_{\st})+\Delta_{L}(z,k_{\st})+\Delta_{T}(z,k_{\st})+\Delta_{LL}(z,k_{\st})+\Delta_{LT}(z,k_{\st})+\Delta_{TT}(z,k_{\st}),
\label{e:fragcor}
\ee
where the $\Delta_{U}(z,k_{\st})$, $\Delta_{L}(z,k_{\st})$ and $\Delta_{T}(z,k_{\st})$ are the correlators that describe fragmentation into an unpolarized, longitudinally and transversely polarized spin 1/2 particle. For spin 0 only $\Delta_{U}(z,k_{\st})$ is relevant. The correlators in Eq.~\ref{e:fragcor} are with the notation of Ref.~\cite{Bacchetta:2006tn} at leading twist given by
\bea
\Delta_{U}(z,k_{\st})&=&\,\bigg\{D_{1}(z,k_\st^2)+iH_{1}^{\perp}(z,k_\st^2)\frac{\slashed{k}_{\st}}{M_h}\bigg\}\frac{\slashed{K}}{2}, \\
\Delta_{L}(z,k_{\st})&=&\,\bigg\{G_{1L}(z,k_\st^2)S_{h\,\sL}\gamma_{5}+H_{1L}^{\perp}(z,k_\st^2)S_{h\,\sL}\frac{\gamma_{5}\,\slashed{k}_{\st}}{M_h}\bigg\}\frac{\slashed{K}}{2}, \\
\Delta_{T}(z,k_{\st})&=&\,\bigg\{-G_{1T}(z,k_\st^2)\frac{k_{\st}\cdot S_{h\,\st}}{M_h}\gamma_{5}+H_{1T}(z,k_\st^2)\gamma_{5}\,\slashed{S}_{h\st} \nonumber \\
&&\hspace{8mm}-\,H_{1T}^{\perp}(z,k_\st^2)\frac{k_{\st}\cdot S_{h\,\st}}{M_h}\frac{\gamma_{5}\,\slashed{k}_{\st}}{M_h}+D_{1T}^{\perp}(z,k_\st^2)\frac{\epsilon_{\st}^{\rho\sigma}k_{\st\rho}S_{h\,\st\sigma}}{M_h}\bigg\}\frac{\slashed{K}}{2},
\eea
\bea
\Delta_{LL}(z,k_{\st})&=&\,\bigg\{D_{1LL}(z,k_\st^2)S_{h\,\sL\sL}+iH_{1LL}^{\perp}(z,k_\st^2)S_{h\,\sL\sL}\frac{\slashed{k}_{\st}}{M_h}\bigg\}\frac{\slashed{K}}{2}, \\
\Delta_{LT}(z,k_{\st})&=&\,\bigg\{-D_{1LT}(z,k_\st^2)\frac{k_{\st}\cdot S_{h\,\sL\st}}{M_h}-G_{1LT}(z,k_\st^2)\epsilon_{\st}^{\mu\nu}S_{h\,\sL\st\,\mu}\frac{k_{\st\,\nu}}{M_h}\gamma_{5} \nonumber \\
&&\hspace{8mm}-H_{1LT}^{\prime}(z,k_\st^2)\gamma_{5}\gamma_{\nu}\epsilon_{\st}^{\nu\rho}S_{h\,\sL\st\,\rho}-iH_{1LT}^{\perp}(z,k_\st^2)\frac{k_{\st}\cdot S_{h\,\sL\st}}{M_h}\frac{\slashed{k}_{\st}}{M_h}\bigg\}\frac{\slashed{K}}{2}, \\
\Delta_{TT}(z,k_{\st})&=&\,\bigg\{D_{1TT}(z,k_\st^2)\frac{k_{\st\,\alpha\beta} S_{h\,\st\st}^{\alpha\beta}}{M_h^2}+G_{1TT}(z,k_\st^2)\epsilon_{\st}^{\mu\nu}S_{h\,\st\st\,\nu\rho}\frac{k_{\st}^{\rho}k_{\st\,\mu}}{M_h^2}\gamma_{5} \nonumber \\
&&\hspace{8mm}+H_{1TT}^{\prime}(z,k_\st^2)\gamma_{5}\gamma_{\nu}\epsilon_{\st}^{\nu\rho}S_{h\,\st\st\,\rho\sigma}\frac{k_{\st}^{\sigma}}{M_h}+iH_{1TT}^{\perp}(z,k_\st^2)\frac{k_{\st\,\alpha\beta} S_{h\,\st\st}^{\alpha\beta}}{M_h^2}\frac{\slashed{k}_{\st}}{M_h}\bigg\}\frac{\slashed{K}}{2},
\eea
where $D_{1T}^{\perp}$, $H_{1}^{\perp}$, $H_{1LL}^{\perp}$, $G_{1LT}$, $H_{1LT}^{\prime}$, $H_{1LT}^{\perp}$, $G_{1TT}$, $H_{1TT}^{\prime}$ and $H_{1TT}^{\perp}$ are T-odd. The relative sign difference between certain corresponding TMD PDF and TMD PFF contributions comes from the definition $\epsilon_{\st}^{\alpha\beta}=\epsilon^{\alpha\beta\rho\sigma}n_{+\rho}n_{-\sigma}$, where interchanging $n_+$ and $n_-$ gives a relative minus sign~\cite{Bacchetta:2006tn}. The TMD PFFs $G_{1}(z,k_\st^2)$, $H_{1}(z,k_\st^2)$, $H_{1LT}(z,k_\st^2)$ and $H_{1TT}(z,k_\st^2)$ are defined in the same way as their TMD PDF counterparts, whereas integrated TMD PFFs are defined as
\be
D_{\ldots}(z)=z^2 \int d^2 k_{\st} D_{\ldots}(z,k_{\st}^{2}).
\ee

\end{document}